\documentclass[aps,prd,reprint,groupedaddress,floatfix]{revtex4-1}
\usepackage{graphicx}
\usepackage{subfigure,epsfig,amsfonts}
\usepackage{amsmath}
\usepackage{amssymb}
\usepackage{amsthm}
\begin{document}

%\preprint{}

\title{Will LISA Detect Harmonic Gravitational Waves from Galactic Cosmic String Loops?}


\author{Zimu Khakhaleva-Li}
\email{zimu@uchicago.edu}
%\thanks{}
\affiliation{Department of Physics, University of Chicago, Chicago, Illinois 60637, USA}
\author{Craig J. Hogan}
\email{craighogan@uchicago.edu}
\affiliation{Department of Astronomy and Astrophysics, University of Chicago, Chicago, Illinois 60637, USA}
\affiliation{Fermi National Accelerator Laboratory, Batavia, Illinois 60510, USA}


\date{\today}

\begin{abstract}
Rapid advancement in the observation of cosmic strings has been made in recent years placing increasingly stringent constraints on their properties, with $G\mu\lesssim 10^{-11}$ from Pulsar Timing Array (PTA) observations. Cosmic string loops with low string tension clump in the halo of the Galaxy due to the combination of slow loop decay and low gravitational recoil, resulting in great enhancement to loop abundance in the Galaxy. With an average separation of down to just a fraction of a kpc, and the total power of gravitational wave (GW) emission dominated by harmonic modes spanning a wide angular scale, resolved loops located in proximity to the solar system are powerful, persistent, and highly monochromatic sources of GW with a harmonic signature not replicated by any other sources, making them prime targets for direct detection by the upcoming Laser Interferometer Space Antenna (LISA), whose frequency range is well-matched for this task. Unlike detection of bursts where the detection rate scales with loop abundance, the detection rate for harmonic signal is the result of a complex interplay between the strength of GW emission by a loop, loop abundance, and other sources of noise, and is most suitably investigated through numerical simulations. We develop a robust and flexible framework for simulating loops in the Galaxy for predicting direct detection of harmonic signal from resolved loops by LISA. Our simulation reveals that the most accessible region in the parameter space for direct detection consists of large loops $\alpha=0.1$ with low string tension $10^{-21}\lesssim G\mu\lesssim 10^{-19}$. Direct detection of field theory cosmic strings is unlikely, with the detection probability $p_{\mathrm{det}}\lesssim 2\%$ for a 1-year mission under the most favorable conditions. An extension of our results suggests that direct detection of cosmic superstrings with a low intercommutation probability is very promising, even unavoidable with optimal parameters. Searching for harmonic GW signal from resolved loops through LISA observations will potentially place physical constraints on string theory.
\end{abstract}

%\keywords{}

\maketitle

\section{Introduction}
Cosmic strings can be created naturally at the end of inflation as effectively one-dimensional (1D) topological defects from spontaneous symmetry breaking (SSB) of a gauge symmetry\cite{VS1994,Ringeval2010}. Such 1D topological defects are not purely theoretical, and e.g., have been produced during phase transition of liquid crystals\cite{LCD}. A network of infinite cosmic strings is then created through the Kibble mechanism\cite{Kibble1976}, with the correlation length-scale $L(t)\lesssim t$ from causality, and string tension
\begin{equation}
\label{eqn:Gmustr}
G\mu\sim\left(\frac{\eta}{m_{\mathrm{pl}}}\right)^2.
\end{equation}
As an interesting coincidence, cosmic strings with $G\mu\sim 10^{-6}$ corresponding to the Grand Unified Theory (GUT) energy scale have the right magnitude to be responsible for the primordial density perturbation\cite{VS1994,strCMB,Ringeval2010,ChernoffBurst}, but do not agree with the observed spectrum of the Cosmic Microwave Background (CMB)\cite{Ringeval2010}. GUT cosmic strings have long been ruled out by various observations\cite{obsCMB1,obsCMB2,Chernoffearly,obslensing2,obslensing3,obslensing4,obsBBN1,obsBBN2,HoganRees,obsPTA,Gulimit,obsVirgo,obsLIGO}. Macroscopic string-like objects can alternatively be created as F-strings and D1-strings at the end of various brane inflation scenarios under the string theory framework\cite{Chernoff2014}, with greatly relaxed theoretical restrictions on their properties due to the richness of the framework.

Unlike undesirable topological defects such as domain walls and monopoles created from SSBs before inflation and then inflated away\cite{inflation1,inflation2,inflation3}, cosmic strings are naturally created at the end of inflation, and can easily be made compatible with observations\cite{Chernoff2014}, making their existence a very real possibility. Instead of being avoided, properties of these natural relics of the early universe should be studied and constrained with observations. Preserving the physics at the SSB energy scale, they serve as a window through which one can get a direct glimpse at physics far beyond the energy scale currently accessible through colliders. Or, they can be macroscopic remnants created directly from string theory. Detecting cosmic strings can therefore open a gateway towards understanding new physics beyond the Standard Model.

Cosmic string loops much smaller than the horizon are created from intercommutation of infinite strings\cite{VS1994}. As 1D infinite objects stretching with the cosmic expansion, their energy density is only diluted by $\rho_{\infty}\propto a(t)^{-2}$, meaning that the total energy grows as $E_{\infty}\propto a(t)$. Infinite strings will therefore quickly dominate the universe without a mechanism for losing energy. Loop formation is the dominant energy loss mechanism\cite{VS1994,Ringeval2007}. As massive objects moving at relativistic velocities\cite{VS1994}, cosmic strings are powerful sources of gravitational wave (GW) in the universe, and detecting their GW emission is therefore an important method for direct observation. Both subhorizon loops and the infinite string network emit GW, with the dominant emission coming from the former\cite{VS1994,Auclair2019} as a result of vibrational modes of loops, with frequencies simply determined by loop sizes.

Detection of the stochastic background of GW from all loops in the universe over all time has been studied extensively, and considerable progress on constraining $G\mu$ has been made recently\cite{Auclair2019}, with the current constraint $G\mu\lesssim 10^{-11}$ from Pulsar Timing Array (PTA) observations\cite{obsPTA,Gulimit}. The recent comprehensive study\cite{Auclair2019} has concluded that the upcoming space-based GW observatory, the Laser Interferometer Space Antenna (LISA)\cite{LISAwhite,LISAsens,LISAwp} can detect the stochastic background down to at least $G\mu\sim 10^{-17}$.

Loops with such low $G\mu$ are long-lived with the potential to survive to the present time without significant loop decay, allowing sufficient time for their originally relativistic peculiar velocity to be damped away by the cosmic expansion\cite{CH09}. Such loops should clump in the Galaxy like cold dark matter (CDM)\cite{Chernoffearly,DP09,CH09}, resulting in a significant enhancement of loop abundance in the Galaxy, estimated to be at least by a factor of $10^5$\cite{Chernoffearly,CH09,ChernoffBurst}. Thus, detecting GW from Galactic loops could be an even more promising approach to observing cosmic strings. Galactic loops emitting GW in the PTA frequency range have lower abundance, since these larger loops are created later in time when the horizon is larger. At the opposite end of the spectrum, unless $G\mu\lesssim 10^{-22}$, loops emitting GW in the frequency range of ground-based observatories such as the Laser Interferometer Gravitational-Wave Observatory (LIGO)\cite{aLIGOsens} have already decayed. Loops with $G\mu\lesssim 10^{-22}$ have very weak signal, presenting a challenge for ground-based observatories. The LISA frequency range and sensitivity are well suited for further extending observational bounds by observing loops clumped in the Galaxy.

While the total unresolved signal from Galactic loops forms the Galactic background which adds to the stochastic background, the more interesting prospect from the high loop abundance in the Galaxy due to clumping is the possibility of direct observation of GW from individual loops, i.e. direct detection of single loops, from which a better understanding of the properties of cosmic strings may be possible. There are two approaches for attempting this. One can try to resolve bursts emitted by small-scale features on loops such as cusps and kinks\cite{VS1994,cuspkey,kinkkey}. Such highly beamed signal only accounts for a small fraction of the total power of GW emission from loops and can be transient. The detection rate in this case scales with loop abundance, and has been estimated theoretically in ref. \cite{ChernoffBurst}.

The alternative approach first proposed by ref. \cite{DP09} but has never been explored in detail before takes advantage of the high loop abundance in the Galaxy in a different way. As a result of clumping, at $G\mu=10^{-20}$, even considering gravitational recoil due to anisotropy in GW emission by loops\cite{CH09}, there can be more than $10^8$ loops in the Galaxy emitting harmonic GW in just one frequency bin of size $\Delta f=1/T_{\mathrm{obs}}\sim 3\times 10^{-8}\ \mathrm{Hz}$ for a 1-year mission, in the sensitive region of LISA, $10^{-3}\ \mathrm{Hz}\lesssim f\lesssim 10^{-1}\ \mathrm{Hz}$. The expected distance from the solar system to the closest such loop is then just a fraction of a kpc, raising the prospect that some loops may be located in extreme proximity that their harmonic GW emission can overwhelm all other sources and be detected directly. Such harmonic signals dominate the total power of GW emission by loops, and are persistent, highly monochromatic, and emitted over a wide angular scale, with the unmistakable harmonic signature differentiating them from all other sources. In the Fourier transform of the LISA strain signal, this harmonic signature will appear as tall spikes towering over the background at frequencies that are exact integer multiples of the fundamental. Such signals are not replicated by any other source, including white dwarf (WD) binaries which are also persistent and monochromatic, making them extremely easy to search for. Loops in the neighborhood of the solar system are therefore highly tantalizing targets for LISA. Unlike detection of bursts, detection of harmonic signals from individual loops relies on them being located sufficiently closely in relation to other loops, and also depends on relative levels of noises. The detection rate is the result of a complex interplay between the strength of GW emission from a loop, loop abundance in the Galaxy, and other sources of noise, and does not scale simply with loop abundance. It is therefore most appropriately analyzed through numerical simulations of loops in the Galaxy, forming the objective of this study.

We organize this paper as follows. After introducing cosmic strings and the goal of our study in this section, in section II, we discuss some theoretical background directly contributing to and essential for comprehending the meaning and significance of this study. We develop the framework and all methodologies for our simulation using toy models in section III, where we also make predictions of expected physical results. Simulation results for the physical Galaxy are presented in section IV. We conclude in section V, and discuss some implications of our results on cosmic superstrings.

\section{Theoretical Background}
\subsection{Cosmic String Loop Formation}
\label{sssec:onescale}
\label{sssec:alphatho}
Motion of the network of infinite strings on superhorizon scales is effectively frozen, stretching with the cosmic expansion, while they move freely and exhibit Brownian motion on subhorizon scales\cite{VS1994}, giving rise to the possibility of collisions. When string segments collide, there is a probability $P_{\mathrm{int}}$ that they intercommute, potentially forming loops. In this work, we adopt canonical Nambu-Goto cosmic strings formed in the field theory framework, which are basically always assumed to have $P_{\mathrm{int}}=1$\cite{VS1994,Auclair2019}. Cosmic superstrings can have much lower $P_{\mathrm{int}}$\cite{Chernoff2014}, we offer some discussions on such models in section \ref{sec:conclude}.

The basic idea that the entire process of loop formation, including the initial loop size, the loop formation rate, and the energy density of the infinite string network, converges to a scaling solution with respect to the horizon originates from Kibble\cite{Kibblea}. This elegant idea of the one-scale model essentially dictates that given physical quantities such as $G\mu$ and $P_{\mathrm{int}}$, there really is just one phenomenological free parameter, the chopping efficiency $c$, encapsulating effects of all processes in the potentially complex series of cosmic string intercommutations before a stable loop is created\cite{Auclair2019}.

With loop formation as the dominant energy loss mechanism\cite{VS1994,Ringeval2007}, the evolution of the infinite string network\cite{Auclair2019} is simply given by,
\begin{equation}
\label{eqn:infstrevo}
\frac{d\rho_{\infty}}{dt}=-2H(t)\left(1+\bar{v}^2\right)\rho_{\infty}-\mu\int_0^{\infty}lf(l,t)dl,
\end{equation}
where $\bar{v}$ is the root-mean-square (rms) velocity, and $f(l,t)$ is the distribution of loop production function. For the one-scale model, $f(l,t)$ is a Dirac delta function, and eq. \ref{eqn:infstrevo} reduces to
\begin{equation}
\label{eqn:onescaleevo}
\frac{d\rho_{\infty}}{dt}=-2H(t)\left(1+\bar{v}^2\right)\rho_{\infty}-c\frac{\rho_{\infty}}{L},
\end{equation}
where $c$ is the chopping efficiency discussed above.

In the one-scale model, the total length of loops created in a Hubble volume over a log time interval scales with the horizon, meaning that in terms of the horizon, the total length is constant. Then the loop number density in log-scale of the loop creation time $t_{\mathrm{c}}$ can be written down naturally with just an overall normalization, which arises from the scaling solution, and is usually obtained from numerical simulations\cite{VS1994,chopeff}. Incorporating effects of dilution due to the cosmic expansion, we have\cite{DP07,DPdiss}
\begin{equation}
\label{eqn:ininumdens}
n(t_{\mathrm{c}},t)=\frac{l_{\mathrm{t}}}{\alpha}H(t_{\mathrm{c}})^3\left(\frac{a(t_{\mathrm{c}})}{a(t)}\right)^3,
\end{equation}
where $\alpha=\frac{l}{t_{\mathrm{c}}}$, $t$ is the time of observation. Consistent with previous studies, we adopt $l_{\mathrm{t}}=8.111$ with $\Delta t_{\mathrm{c}}=0.1t_{\mathrm{c}}$\cite{DP07,DPdiss,Hogan2006,CaldwellAllen}.

Various attempts have been made in the literature to deduce the scaling solution of $\alpha$ from numerical simulations. Very early simulations usually produced very small loops bound by simulation resolutions\cite{VS1994,earlya1,earlya2,earlya3}. Such resolution problems were overcome in refs. \cite{bigloopsim0,bigloopsim1,bigloopsim2}, obtaining a physically meaningful loop production function. They conclude that loop formation does indeed converge to scaling solutions, and is dominated by large loops with $\alpha\sim 0.1$, accounting for most of the length of loops created. Loop formation is strongly suppressed below the gravitational backreaction length-scale, $\alpha\lesssim\frac{l_{\mathrm{br}}}{t}\sim\Gamma G\mu$\cite{MartinsShellard2006,Ringeval2007}, $\Gamma\sim 50$\cite{DP07,Auclair2019}. We call these models the big-loop paradigm. A radically different approach is taken in refs. \cite{smallloop1,smallloop2,smallloop3} which theoretically derive the loop production function, where $l_{\mathrm{br}}$ is also reduced to $l_{\mathrm{br}}\sim 20(G\mu)^{1+2\chi}$, with $\chi_{\mathrm{r}}=0.200^{+0.07}_{-0.10}$ and $\chi_{\mathrm{m}}=0.295^{+0.03}_{-0.04}$ for radiation and matter eras, respectively\cite{smallloopsim,smallloopsto}. The loop formation simulation based on these results\cite{smallloopsim} predicts that loop production is not cut off by gravitational backreaction, resulting in a great overabundance of very tiny loops. We call this model the small-loop paradigm.

Recent loop formation simulations\cite{bigloopnew1,bigloopnew2} still follow the big-loop paradigm. Here, the difference between loops modeled by the one-scale model and those modeled by a full distribution of $\alpha$ is not very significant for GW detection\cite{DP07,Auclair2019}. In particular, any difference mostly manifests in the high frequency flat region of the stochastic background, which for $G\mu$ of interest in this study as we will discuss in section \ref{sssec:summary}, exists outside of the LISA frequency range. The standard one-scale model adopted by studies on GW detection has $\alpha\sim 0.1$\cite{DP07,DP09,ChernoffBurst,Auclair2019}. For such studies, the most important factor is not the distribution of $\alpha$, but the fact that the scaling solution has a cutoff at $\alpha\sim 0.1$, because these largest loops essentially represent the best-case scenario for detection\cite{Auclair2019}. We therefore adopt the one-scale model with $\alpha=0.1$ and $\alpha=10^{-5}$ for our simulation. Results from the latter can help shed light on models with small loops.

\subsection{Gravitational Radiation from Cosmic String Loops}
\label{sssec:loopdyna}
\label{sssec:GWspectho}
As cosmic strings are effectively 1D objects with linear mass density $\mu$, GW is emitted from movements of string segments. The dynamics of a loop is governed by the Nambu-Goto action\cite{VS1994}
\begin{equation}
\label{eqn:NGact}
S=-\mu\int d^2\sigma\sqrt{-\gamma},
\end{equation}
where $\gamma$ is the determinant of the worldsheet metric. When the background spacetime is essentially flat, i.e. $\frac{l}{\mathcal{R}}\sim 0$, where $\mathcal{R}$ is the scale of curvature, loop trajectories are described by simple and intuitive equations of motion. Motion along the loop is relativistic with $\bar{v}^2=0.5$, and the frequency of the fundamental mode of oscillations
\begin{equation}
\label{eqn:f0}
f=\frac{2}{l}.
\end{equation}
Tangent vectors of left- and right-moving loop trajectories are constrained on a unit sphere $\mathbf{X}_{\pm}'^2=1$ called the Kibble-Turok (KT) sphere\cite{traject1}, and loop motion becomes luminal when they intersect, forming a cusp. Solutions require $\mathbf{X}_{\pm}'$ to occupy both hemispheres of the KT sphere, making intersections rather difficult to avoid, which means that cusps are generic features of loops\cite{VS1994}. Another interesting feature consists of kinks which are discontinuities along $\mathbf{X}_{\pm}'$. Intersections are easier to avoid with the presence of discontinuities, meaning that kinks, especially a large number of them, will tend to inhibit cusps. This has been confirmed both theoretically\cite{kinkearly,kinkkey,smallloopsto} and through numerical simulations\cite{krcsim}. Though kinks are expected to be produced during string intercommutation\cite{kinkearly,VS1994,smallloopsto}, they are smoothed out relatively quickly\cite{kinksmooth} at a timescale $t_{\mathrm{kink}}\sim\frac{l}{n_{\mathrm{k}}\Gamma G\mu}$, where $n_{\mathrm{k}}=\frac{l}{l_{\mathrm{kink}}}$, which can be much shorter than the loop decay timescale $t_{\mathrm{d}}\sim\frac{l}{\Gamma G\mu}$. Cusps are therefore generally expected to dominate\cite{DP07,BlancoOlum2017}. However, some recent numerical simulations suggest that cusps may not dominate over kinks when the latter are present in large numbers\cite{kinkdom1,kinkdom2}. We therefore consider both cases in our simulation.

Cusps and kinks are associated with highly relativistic motion along loops, and produce bursts of powerful gravitational radiation at frequencies much higher than the fundamental\cite{kinkkey,DP07,ChernoffBurst}. As general features of loops, their constant presence effectively modifies the high frequency spectrum, from a typical quadrupole spectrum to power-laws\cite{VS1994}.

Finding solutions for loop trajectories analytically is very difficult. The simplest nontrivial loop trajectory involving the 1st and 3rd modes has been found\cite{traject1,traject2}, and contains cusps. The 2nd mode is excluded mathematically\cite{no2har}. With a solution for the loop trajectory, the energy-momentum tensor of the loop takes on a simple form in transverse gauge\cite{cuspkey},
\begin{equation}
\label{eqn:Tuv}
T^{\mu\nu}(t,\mathbf{X})=\mu\int d\sigma\left(\dot{x}^{\mu}\dot{x}^{\nu}-x'^{\mu}x'^{\nu}\right)\delta^{(3)}(\mathbf{X}-\mathbf{X}(t,\sigma)).
\end{equation}
The power of GW in the frequency mode per solid angle can be calculated using the Fourier transform $T^{\mu\nu}(f_n,\mathbf{k})$\cite{Weinberg1972},
\begin{equation}
\label{eqn:PnT}
\frac{d\dot{E}_n}{d\Omega}=4\pi Gf_n^2\left(T^*_{\mu\nu}(f_n,\mathbf{k})T^{\mu\nu}(f_n,\mathbf{k})-\frac{1}{2}\left|T_{\nu}^{\nu}(f_n,\mathbf{k})\right|^2\right).
\end{equation}
The total power is then\cite{VS1994,DP07},
\begin{equation}
\label{eqn:Ptot}
P=\sum_{n=1}^{\infty}\int d\Omega\frac{d\dot{E}_n}{d\Omega}\equiv\Gamma G\mu^2\equiv\sum_{n=1}^{\infty}P_nG\mu^2,
\end{equation}
where the dimensionless parameters $\Gamma$ and $P_n$ parametrize the total power and the power in each frequency mode, respectively, with
\begin{equation}
\label{eqn:Gammanorm}
\Gamma=\sum_{n=1}^{\infty}P_n.
\end{equation}
For power-law GW emission spectra,
\begin{equation}
\label{eqn:Pnnorm}
P_n=\frac{\Gamma}{\zeta(s)}n^{-s},
\end{equation}
where $\zeta(s)$ is the Riemann zeta function. Loops decay at the timescale\cite{VS1994}
\begin{equation}
\label{eqn:decayt}
t_{\mathrm{d}}\sim\frac{E}{P}=\frac{l\mu}{\Gamma G\mu^2}=\frac{l}{\Gamma G\mu}.
\end{equation}
Then, at $t$, the size of a loop created at $t_{\mathrm{c}}$ is\cite{DP07}
\begin{equation}
\label{eqn:sizet}
l(t_{\mathrm{c}},t)=\alpha t_{\mathrm{c}}-\Gamma G\mu(t-t_{\mathrm{c}}).
\end{equation}
This allows us to rewrite eq. \ref{eqn:ininumdens} in terms of the frequency of GW emission at time $t$ as $n(f,t)$. Loops decay slowly compared to the periods of their GW emission, meaning that frequency modes should be very well-defined. We can see this from eq. \ref{eqn:Ptot}\cite{DP09},
\begin{equation}
Q_n=2\pi f_n\frac{E}{\dot{E}_n}=\frac{4\pi n}{l}\frac{\mu l}{P_n G\mu^2}=\frac{4\pi n}{P_n G\mu}.
\end{equation}

The parameter $\Gamma$ is independent of loop sizes, but does depend on loop trajectories\cite{VS1994}. Early derivations and numerical simulations produced a range of values, $44\lesssim\Gamma\lesssim 100$\cite{VS1994,AllenCasper,bigloopnew2}. More recent studies tend to converge at $\Gamma\sim 50$\cite{kinkkey,DP07,DP09,DPdiss,BBHS,ChernoffBurst,Auclair2019}, for both the cusps- and the kinks-dominated spectra. The range of possible $\Gamma$ means that a detailed estimate of this parameter is not important for our purpose.

The cusps-dominated GW emission spectrum takes on the form
\begin{equation}
\label{eqn:cuspspec}
P_n\propto n^{-\frac{4}{3}},
\end{equation}
over an angular scale
\begin{equation}
\label{eqn:cuspangle}
\theta_n\sim n^{-\frac{1}{3}},
\end{equation}
from both analytical and numerical studies of loop trajectories\cite{VS1994,specsup1,gen13sol,BBHS,smallloopsto,cuspsim,BlancoOlum2017}. Mathematical pursuits of more general solutions to loop trajectories continue\cite{Anderson2003}. However, higher oscillation modes are likely unimportant physically, as they are expected to be damped by the cosmic expansion soon after formation\cite{VS1994}. Since cusps usually dominate, eq. \ref{eqn:cuspspec} is frequently adopted by studies of GW from loops\cite{DP07,DP09,DPdiss,CH09,cuspsupp1,cuspsupp2,bigloopnew2,ChernoffBurst}. Early calculations of kinks-dominated trajectories indicated a steeper spectrum $P_n\propto n^{-2}$\cite{kinkearly,VS1994,specsup1}. More recent calculations conclude that the kinks-dominated spectrum has the form\cite{kinkkey,BBHS}
\begin{equation}
\label{eqn:kinkspec}
P_n\propto n^{-\frac{5}{3}},
\end{equation}
with anisotropy in emission only applying to one angular direction, giving rise to a ``fan-like'' pattern. Compared with eq. \ref{eqn:cuspspec}, we see that GW from cusps tends to dominate over that from kinks when both are present\cite{smallloopsto}.

From eq. \ref{eqn:cuspangle} and the discussion above, we see that though not exactly isotropic, GW from loops is beamed only for high frequency burst modes. Our study focuses on low frequency harmonic modes, where the power of emission is distributed over a large solid angle. Since both GW emission spectra are relatively steep, bursts account for a very small fraction of the total power of radiation, with eq. \ref{eqn:Gammanorm} showing that close to half of the total power emitted goes into the fundamental mode alone. For a given frequency range, burst signals are emitted by much larger loops, which are created later and therefore have lower number densities than those of loops emitting harmonic signals. There also exists the possibility that the orientation of a loop can change over time, making burst signals transient.

\subsection{Cosmic String Loops in the Galaxy}
\label{sssec:clumptho}
\label{sssec:rockettho}
Cosmic strings in the original picture thought to have high $G\mu$ corresponding to the GUT scale lose energy rapidly through gravitational radiation and decay away quickly. The primary effect of such loops on the galactic scale is to possibly seed structure formation through their density perturbation\cite{VS1994,CH09}. In the modern picture of cosmic strings with very low $G\mu$, loops decay slowly and can survive for many Hubble times, and their peculiar velocities are damped by the cosmic expansion\cite{VS1994},
\begin{equation}
\label{eqn:pdamp}
p\propto a^{-1},
\end{equation}
where $p$ is the momentum. The peculiar velocities of these long-lived loops created in the radiation era will have been completely non-relativistic for a long time by now. They should therefore behave similarly to CDM on the galactic scale, and clump in the halo of the Galaxy following the CDM density profile, resulting in a great overdensity of loops in the Galaxy. This idea is first proposed by refs. \cite{Chernoffearly,DP09}, and the corresponding overdensity is estimated to be at least a factor of $10^5$.

A detailed analysis of loop clumping in the Galaxy is carried out in ref. \cite{CH09}. Since it takes many Hubble times to sufficiently damp the peculiar velocity of a loop so that it becomes non-relativistic, loop clumping favors loops with smaller $t_{\mathrm{c}}$, i.e. loops with larger $\alpha$ and lower $G\mu$. Ref. \cite{CH09} finds that light loops with $\alpha=0.1$ clump very well following the CDM density profile, with clumping becoming essentially complete for $G\mu\lesssim 10^{-13}$. Loops from the small-loop paradigm with $\alpha=20(G\mu)^{1+2\chi}$ on the other hand, experience little clumping even with very low $G\mu$, making them irrelevant for this study. We therefore assume complete clumping in our study, which is somewhat optimistic for loops with $\alpha=10^{-5}$, if they give any detection at all that is.

From the power of GW emission by a loop in eq. \ref{eqn:Ptot}, the flux received by the detector is
\begin{equation}
\label{eqn:fluxind}
F(r')=\frac{\Gamma G\mu^2}{4\pi r'^2},
\end{equation}
where $r'$ is from the solar system, and we always treat the emission as isotropic, as low frequency harmonic modes have wide angular scales. This can be integrated to obtain the total flux from all loops in the Galaxy,
\begin{equation}
\label{eqn:fluxtot}
F_{\mathrm{g}}(f)=\int F(r')n_{\mathrm{g}}(r,f)dV,
\end{equation}
where $n_{\mathrm{g}}(r,f)$ is the loop number density in the Galaxy, $r$ is from the Galactic center. Low frequency harmonic modes of GW from loops are well represented by plane-wave solutions of linearized gravity\cite{kinkkey,DP09}, with the maximum wavelength in the LISA frequency range less than solar system size, making it appropriate to adopt the short wavelength approximation. In the transverse traceless gauge with harmonic gauge condition, the Isaacson stress-energy tensor simplifies to\cite{hloop1,hloop2}
\begin{equation}
\label{eqn:IsaT}
T_{\mu\nu}=\frac{1}{32\pi G}\left\langle\frac{\partial h_{ij}}{\partial x^{\mu}}\frac{\partial h^{ij}}{\partial x^{\nu}}\right\rangle.
\end{equation}
Substituting in the plane-wave solution, the flux is\cite{hloop2}
\begin{equation}
\label{eqn:fluxcomp}
F=T^{03}=\frac{\pi f^2}{8G}\langle h_{ij}h^{ij}\rangle=\frac{\pi f^2 h^2}{8G},
\end{equation}
where $h$ is the strain amplitude. For harmonic modes of a loop,
\begin{equation}
\label{eqn:hindloop}
h_n=\frac{1}{f_n}\sqrt{\frac{8GF_n}{\pi}}=\frac{\sqrt{2P_n}G\mu}{\pi rf_n}.
\end{equation}

\begin{figure}[htb]
\includegraphics[width=\columnwidth]{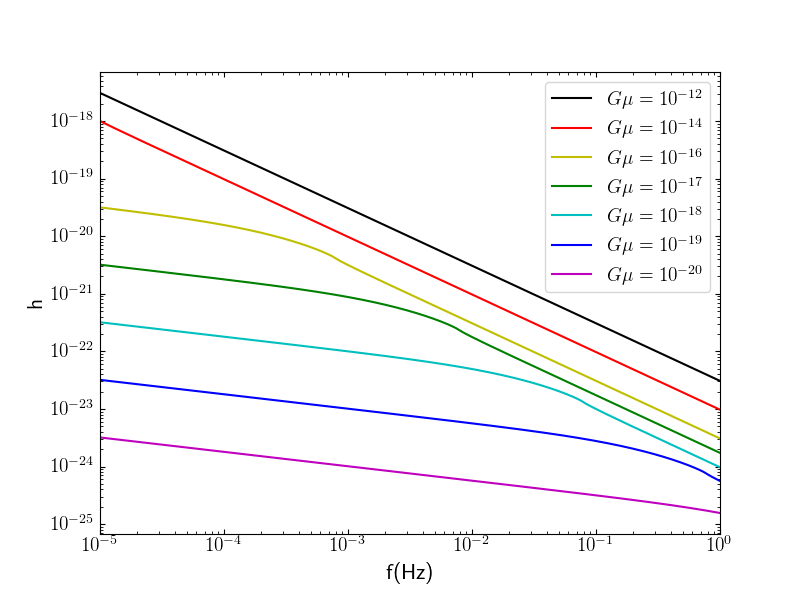}
\caption{The total strain amplitude from all loops in the Galaxy in the toy models described in section \ref{sec:simul}, with $\alpha=0.1$ and the rocket effect.}
\label{fig:galhtho}
\end{figure}

We present in fig. \ref{fig:galhtho} the total strain amplitude from all loops in the Galaxy with $\alpha=0.1$ and the rocket effect, which will be discussed below. The change in slope is partially due to the transition into the higher frequency loop decay regime, and the steepness of the slope in this region is also partially due to the rocket effect, which is responsible for the break in the transition. This Galactic background is a confusion noise in our study. We should stress that this plot should not be compared directly with the LISA sensitivity curve, because it is binned according to the log bins of eq. \ref{eqn:ininumdens}. It represents the Galactic background in the toy models described in section \ref{sec:simul}, and the background for the physical Galaxy is presented in fig. \ref{fig:pgalhml}.

There is a competing physical process on the loop peculiar velocity to damping by the cosmic expansion, which is a result of gravitational recoil due to anisotropy in GW emission, appropriately named the rocket effect. This has been analyzed in detail in ref. \cite{CH09} for cusps-dominated loops with fixed overall directions of recoil\cite{CH09,ChernoffBurst}, which can be expressed as the constant force per unit mass
\begin{equation}
\label{eqn:arocket}
a_{\mathrm{r}}=\frac{\Gamma_P G\mu}{l},
\end{equation}
where $\Gamma_P\sim 10$ parametrizes the total momentum carried by GW. For loops with low $G\mu$, the rocket effect slightly reduces clumping at very large radii, which is unimportant for our purpose, as the contribution from these distant regions with low loop abundance is negligible. The more important consequence of the rocket effect is loop ejection from the Galaxy, which is an example of the Stark problem, the classical counterpart to the Stark effect. Complete solutions\cite{Stark1,Stark2,Stark3} have been found indicating that unlike the quantum behavior where stable bound states can exist, classically, the orbit of the object always becomes parabolic in the limit of $t\rightarrow\infty$\cite{Stark1}. Ref. \cite{CH09} derives that clumped loops outside the truncation radius
\begin{equation}
\label{eqn:rtr}
r_{\mathrm{tr}}=r_{\mathrm{ta}}\left(\frac{1.575H_0r_{\mathrm{ta}}\alpha t_{\mathrm{c}}}{\Gamma_P G\mu t_0}\right)^{\frac{4}{5}}
\end{equation}
is ejected from the Galaxy before the present time. Clearly, $\lim_{t_0\rightarrow\infty}r_{\mathrm{tr}}=0$, consistent with the general solution. Even though $\alpha$ appears in the expression, $r_{\mathrm{tr}}$ does not actually depend on $\alpha$ for a given loop size, i.e. frequency of GW emission, since $\alpha t_{\mathrm{c}}$ stays constant.

The rocket effect ejects loops over many orbits, with older loops having smaller $r_{\mathrm{tr}}$, which, for a given $\alpha$, are smaller loops. Truncation is therefore more severe at higher frequency, making $r_{\mathrm{tr}}$ a function of frequency. We can then define a truncation frequency $f_{\odot}$ such that
\begin{equation}
\label{eqn:ftr}
r_{\mathrm{tr}}(f_{\odot})\equiv r_{\odot},
\end{equation}
where $r_{\odot}\sim 8\ \mathrm{kpc}$\cite{DP09}. When $r_{\mathrm{tr}}<r_{\odot}$, there simply cannot exist nearby loops, thereby severely restricting detection of harmonic signal from an individual loop for $f>f_{\odot}$. When $f\lesssim f_{\odot}$, $r_{\mathrm{tr}}\gtrsim r_{\odot}$, and detection can be enhanced because the Galactic background is reduced while loops are still permitted to be located nearby. Loops with higher $G\mu$ experience stronger recoil, and therefore have smaller $r_{\mathrm{tr}}$, meaning that as $G\mu$ increases, $f_{\odot}$ decreases. Comparing eq. \ref{eqn:rtr} with eq. \ref{eqn:sizet}, the relations between $\alpha t_{\mathrm{c}}$ and $\Gamma G\mu t_0$ (or $\Gamma_P G\mu t_0$) mean that the rocket effect has a similar dependence on loop age to that of loop decay. In fact, as the frequency of GW emission increases, it happens that for the solar system, the rocket effect always becomes significant at about the frequency where such loops enter the decay regime.

\begin{figure}[htb]
\includegraphics[width=\columnwidth]{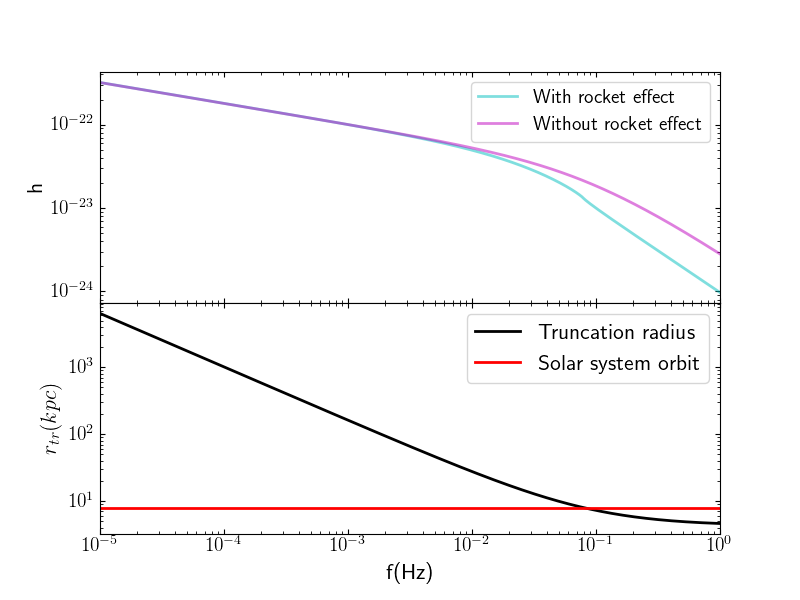}
\caption{The Galactic background in the toy models with $\alpha=0.1$ and $G\mu=10^{-18}$ is plotted in the upper panel, both with (cyan) and without (magenta) the rocket effect. The corresponding $r_{\mathrm{tr}}$ is plotted in the lower panel.}
\label{fig:rcritillu}
\end{figure}

We present an illustration of the truncation of loops in the Galaxy due to the rocket effect in fig. \ref{fig:rcritillu}. The cyan curve in the upper panel is the same Galactic background with $\alpha=0.1$ and $G\mu=10^{-18}$ found in fig. \ref{fig:galhtho}. The upper panel shows that indeed background reduction begins in advance of $f_{\odot}$ as frequency increases. When compared to $r_{\mathrm{tr}}$ plotted in the lower panel, we see that the Galactic background displays the break we mentioned previously at exactly $f_{\odot}$, representing the qualitative change in loop distribution around the solar system when $r_{\mathrm{tr}}$ falls below $r_{\odot}$.

\subsection{LISA Orbital Modulations}
\label{sssec:orbmod}
The LISA spacecraft is an equilateral triangle tilted at $60^{\circ}$ with respect to the ecliptic formed by three orbiters. Each orbiter is in a separate Earth-trailing heliocentric orbit, resulting in an overall full rotation of the triangle over a complete 1-year orbit. The unit vector of the plane of the triangle projected onto the ecliptic always points towards the Sun, meaning that its sweep forms a cone with a $60^{\circ}$ opening over an orbit\cite{LISAwp, LISAwhite}.

The detector patterns of LISA themselves are simple, differing from those of Michelson-type interferometers by only a very intuitive factor of $\frac{\sqrt{3}}{2}$\cite{LISAresp1, LISAresp2}. However, as a space-based detector, the orbital motion of LISA introduces major complications to its detector response. In particular, it induces three types of modulations on the detected signal\cite{orbmotiv, orbmod}:
\begin{enumerate}
\item Amplitude modulations due to the sweep of antenna patterns of LISA;
\item Frequency modulations due to Doppler shifts in the signal caused by the relative motion with respect to the source;
\item Phase modulations due to different detector responses of LISA to the two polarization eigenstates of GW.
\end{enumerate}
All signal modulations above are periodic over a 1-year orbit.

Tracking the LISA orbital motion and associated signal modulations in real time in the simulation is neither possible given current computational capabilities, nor necessary since the simulation makes predictions for a 1-year mission, needing only the total effect averaged over the orbit. This has been calculated by ref. \cite{orbmod} for monochromatic sources, with the total corrections fully specified given source locations, inclinations $\iota$, and polarization angles $\psi$.

\section{The Simulation Developed with Toy Models}
\label{sec:simul}
We develop the methodology necessary to successfully perform simulations predicting direct detection by LISA of harmonic GW from resolved cosmic string loops with toy models of loop number density. Computationally, these toy models come naturally by directly evaluating eq. \ref{eqn:ininumdens} at the LISA frequency resolution $\Delta f=\frac{1}{T_{\mathrm{obs}}}$. Such toy models correspond to hypothetical galaxies with a great overabundance of loops compared to ours, by several orders of magnitude for higher frequency.

While the inflated loop abundance in the toy models certainly presents a challenge for numerical simulations, the use of toy models brings four important advantages.
\begin{enumerate}
\item The loop number density is well-behaved and easy to understand at all $G\mu$ and frequency in the toy models, providing a well-defined platform for developing robust and adaptable simulation methodologies.
\item Detection is enhanced in the toy models, enabling a better understanding of the behavior of results in the parameter space.
\item Results from the toy models provide guidance on promising regions of the parameter space to focus computational resources on for further analysis.
\item Estimates of expected physical results can be obtained from the toy models, as they are still based on the physical Galaxy.
\end{enumerate}

\subsection{Simulation Configuration}
\subsubsection{Cosmological and Astrophysical Parameters}
\label{sssec:params}
For the toy models, we adopt the $\mathrm{\Lambda CDM}$ cosmology with Planck 2015 cosmological parameters\cite{Planck2015}:
\begin{eqnarray}
H_0 & = & 67.74\mathrm{km/s/Mpc}, \\
\Omega_{\mathrm{m,0}} & = & 0.3075, \\
\Omega_{\mathrm{\Lambda,0}} & = & 0.6910, \\
\Omega_{\mathrm{\gamma,0}} & = & 5.389\times 10^{-5}, \\
\label{eqn:omeganu0}
\Omega_{\mathrm{\nu,0}} & = & 1.436\times 10^{-3}.
\end{eqnarray}
Neutrinos are regarded as massive, and $\Omega_{\mathrm{\nu,0}}$ is estimated by refs. \cite{astropy1,astropy2} using the method described in ref. \cite{Komatsu2011}. Due to considerations of computational performance, they are treated as relativistic particles contributing to the radiation density, resulting in a further overall enhancement of the loop number density by $\sim 10$ compared to treating them as massless. This has no negative impact on the toy models. Physical results computed with massless neutrinos are presented in section \ref{ssec:masslessneu}.

The distribution of CDM in the Galaxy is well modeled by the Navarro-Frenk-White (NFW) profile\cite{GNFW1,GNFW2}. We therefore adopt the NFW profile\cite{NFW} for the Galactic dark matter halo, which also models the number density of loops clumped in the Galaxy. Consistent with ref. \cite{DP09}, we adopt $R_s=21.5\ \mathrm{kpc}$ and $C=10$, with the common definition\cite{R2001,R2002,R2003} of $R_{\mathrm{vir}}\sim R_{200}$. The turnaround radius $R_{\mathrm{ta}}=1.1\ \mathrm{Mpc}$ is adopted from ref. \cite{CH09}.

\subsubsection{Gravitational Wave Interference}
With plane wave solutions of linearized gravity in the short wavelength approximation, the total signal as a result of superposition of GW from all loops in the Galaxy in the frequency bin $f$ is simply
\begin{equation}
h(t)=h\sin\left(2\pi ft+\delta\right)
\end{equation}
by the harmonic addition theorem\cite{haraddtho}. Define $A\equiv \sum_{i=1}^N h_i\sin\delta_i$ and $B\equiv \sum_{i=1}^N h_i\cos\delta_i$, where $h_i$ given by eq. \ref{eqn:hindloop} and $\delta_i\in\left(-\pi, \pi\right]$ are individual strain amplitudes and uncorrelated phases, respectively, and $N$ is the total number of loops in the bin. The total strain amplitude is
\begin{equation}
\label{eqn:htot}
h=\sqrt{A^2+B^2},
\end{equation}
with the total phase
\begin{equation}
\label{eqn:deltatot}
\delta=\tan^{-1}\frac{A}{B}+
\begin{cases}
0, & B\geq 0; \\
\pi, & B<0\ \mathrm{and}\ A>0; \\
-\pi, & B<0\ \mathrm{and}\ A\leq 0.
\end{cases}
\end{equation}
The simulation computes the superposition for each frequency bin, as the bins are independent of each other.

\subsubsection{The Closest Loops Method}
\label{sssec:clm}
\begin{figure}[tbh]
\includegraphics[width=\columnwidth]{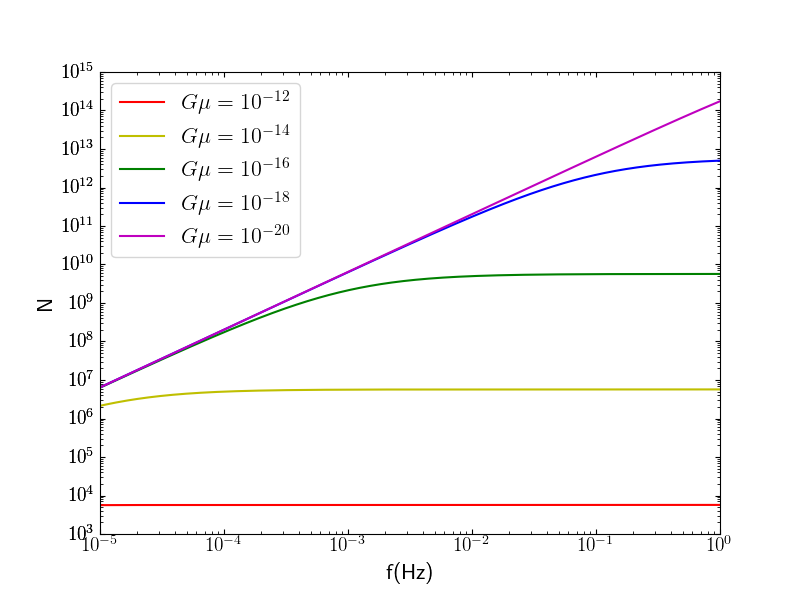}
\caption{The number of loops in the Galaxy in each frequency bin with $\alpha=0.1$ and without the rocket effect.}
\label{fig:Nnr}
\end{figure}

When loop decay is not significant, $N(f)\propto f^{\frac{3}{2}}$ from cosmology. From fig. \ref{fig:Nnr}, this is generally the case at lower frequency, where loops are bigger and are created later, and therefore have not yet experienced significant decay. Loops with lower $G\mu$ experience slower decay and therefore enter the decay regime at higher frequencies. For the LISA frequency range, the decay regime is entirely absent for $G\mu\lesssim 10^{-20}$. The flat region represents the decay regime. For $G\mu\gtrsim 10^{-14}$, loops emitting GW in the entire frequency range have decayed. Loops in the decay regime have similar $t_{\mathrm{c}}$, and therefore similar number densities and sizes initially. However, fractional size differences are amplified by loop decay over time, resulting in these loops having very different sizes today, while their number densities remain similar. In the decay regime, loops with higher $G\mu$ are created much later in time, and hence the lower number densities.

The important point from fig. \ref{fig:Nnr} is that loops can be extremely abundant in the Galaxy, especially at lower $G\mu$. This is clearly not only true in the toy models, but is also the case in the physical model, as rebinning decreases $N(f)$ by a factor of $\frac{1}{T_{\mathrm{obs}}}/0.1f\lesssim 3\times 10^{-7}$. Tracking all loops in the entire Galaxy is certainly far beyond current computational capabilities.

The solution is to only track loops closest to the solar system. For our purpose, it is then possible to recover the statistics for the entire Galaxy based on the outcome of the limited-scale simulation. We call this the ``closest loops method''. We verify this method both theoretically (below) and through simulations (section \ref{sssec:hconf}).

Our simulations have a signal side focusing on accounting for signal from individual nearby loops, and a noise side focusing on generating GW from unresolved background loops in the Galaxy, i.e. the Galactic confusion. The latter at a given frequency is the result of superposition of GW emitted by all loops in the bin. This can vary significantly across frequency bins due to the uncorrelated phases of GW from background loops, even though the theoretical Galactic background is a slowly varying function of frequency as shown in fig. \ref{fig:galhtho}. This variation in the Galactic confusion can mimic the signal to a certain extent. Thus, there are two components to the Galactic confusion for which our simulations need to account for, the Galactic confusion background and the Galactic confusion noise. The former is the mean of the Galactic confusion, $\bar{h}$, which is essentially the theoretically calculated Galactic background. The latter, $h_{\mathrm{n,c}}$, is the amount of variation in the Galactic background across frequency bins.

On the signal side, we will see in section \ref{sssec:decetcomp} that the likelihood of detection at a given frequency, even in the toy models, is so low that most detections are attributed to just the single closest loop. Thus, simulating just a few closest loops at each frequency is sufficient to account for the signal.

On the noise side, we have both $\bar{h}$ and $h_{\mathrm{n,c}}$. Since overall the power of GW emission from background loops adds. Then from eq. \ref{eqn:hindloop},
\begin{equation}
\label{eqn:hbarN}
\bar{h}\propto\sqrt{N}.
\end{equation}
As $h_{\mathrm{n,c}}$ is a result of the superposition of GW from background loops with uncorrelated phases,
\begin{equation}
\label{eqn:hncN}
h_{\mathrm{n,c}}\propto\sqrt{N},
\end{equation}
because the interference is effectively a 1D random walk on the strain amplitude.

Both $\bar{h}$ and $h_{\mathrm{n,c}}$ are modulated by the geometry of the spatial distribution of background loops in the Galaxy. Geometry then cancels when comparing eqs. \ref{eqn:hbarN} and \ref{eqn:hncN}, and we simply have
\begin{equation}
\label{eqn:clm}
\bar{h}\propto h_{\mathrm{n,c}}.
\end{equation}
As $\bar{h}$ is known through theoretical calculations, eq. \ref{eqn:clm} enables us to recover $h_{\mathrm{n,c}}$ for the entire Galaxy by tracking only the closest subset of background loops. The closest loops method proves to be especially powerful for the interesting region of the parameter space where loop abundance is very high.

\subsubsection{Simulation Setup}
\label{sssec:toysetup}
We vary both $G\mu$ and $\alpha$ with and without the rocket effect to cover the parameter space. Simulations are performed with $\alpha=0.1$ and $10^{-5}$, with $G\mu$ varying from $10^{-12}$ down to $10^{-20}$, producing 24 parameter combinations.

Simulations cover the peak LISA sensitivity frequency range $f\in\left[10^{-5}, 1\right]\ \mathrm{Hz}$, with the frequency resolution $\Delta f=\frac{1}{T_{\mathrm{obs}}}$ corresponding to $T_{\mathrm{obs}}=1\ \mathrm{month}$ to conserve computational resources. For the signal side, we run 100 realizations for each simulation set including the closest 10 loops at each frequency to generate detection candidates. For the noise side, we run 10 realizations for each set including the closest 1000 loops at each frequency to estimate the Galactic confusion utilizing the closest loops method.

Additionally, we run 100 signal realizations with the rocket effect at $G\mu=10^{-18}$ and $\alpha=0.1$, with $\Delta f$ corresponding to $T_{\mathrm{obs}}=1\ \mathrm{year}$. This simulation set effectively constitutes a different toy model from the 24 sets above. It helps clarify physical implications of simulating the toy models, and also serves as a consistency test when we renormalize results from the toy models in section \ref{sssec:impphys}.

We consider three main sources of noise:
\begin{enumerate}
\item The Galactic confusion;
\item The stochastic background;
\item The LISA instrumental noise with the Galactic WD binaries background.
\end{enumerate}
We set $T_{\mathrm{obs}}=1\ \mathrm{year}$ for the noise level regardless of the $\Delta f$ that a given set employs, as our goal is to make predictions for the possibility of detection over a 1-year mission period.

The full effects of signal modulations due to the LISA orbital motion averaged over a complete orbit are incorporated, by generating the location of each simulated loop in the Galaxy $(r,\theta,\phi)$ according to the distribution of loop number density in the Galaxy, with random $\iota\in [0,\pi]$ and $\psi\in [0,\pi]$.

\subsection{Data Analysis}
\subsubsection{The Galactic Confusion}
\label{sssec:hconf}
\begin{figure}[htb]
\includegraphics[width=\columnwidth]{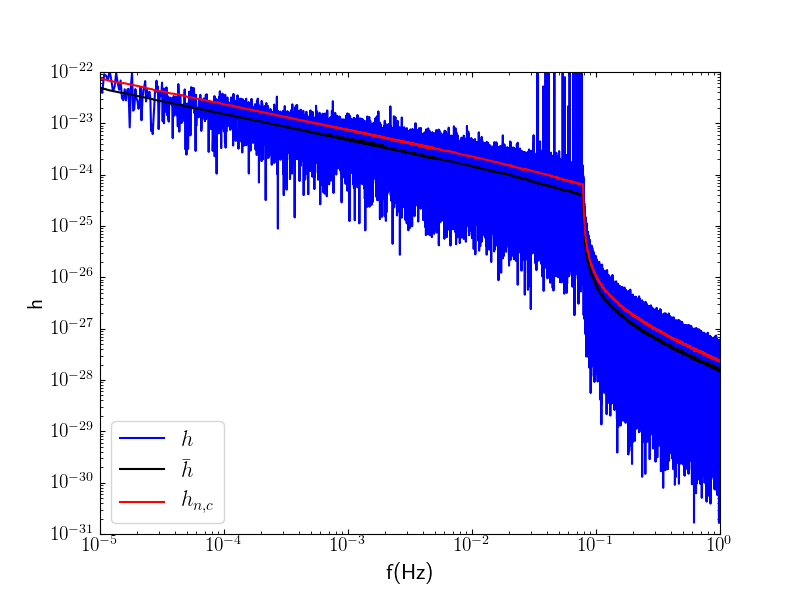}
\caption{Snapshot of a noise side realization with the rocket effect at $G\mu=10^{-18}$ and $\alpha=0.1$. The black and red lines are the directly computed $\bar{h}$ and $h_{\mathrm{n,c}}$, respectively.}
\label{fig:hncexp}
\end{figure}

The cliff-like feature at $f\sim 0.1\ \mathrm{Hz}$ in fig. \ref{fig:hncexp} is located at $f_{\odot}$, and is the same break barely visible in fig. \ref{fig:galhtho} due to the rocket effect. With the absence of the majority of loops in the Galaxy, the feature becomes very pronounced here because the entire population of loops is effectively shifted further away for $f>f_{\odot}$, where distances to these loops become bound by $r_{\mathrm{tr}}$.

The tall vertical lines in $0.01\ \mathrm{Hz}\lesssim f\lesssim 0.1\ \mathrm{Hz}$ in fig. \ref{fig:hncexp} are in fact signal from resolved loops. As they can be several orders of magnitude stronger than $h_{\mathrm{n,c}}$, they need to be carefully excluded on the noise side to avoid skewing estimates of the Galactic confusion. As $\bar{h}$ is a slowly varying function of frequency, we compute $\bar{h}$ and $h_{\mathrm{n,c}}$ at a particular frequency bin by considering the neighborhood of $1\%$ of the total number of frequency bins. The quantities are then estimated from Gaussian fits to the histograms of the neighborhoods. This procedure eliminates the effects of signals located in the far-off tails of the histograms. The estimated $h_{\mathrm{n,c}}$ may appear low visually because of the great abundance of frequency bins at higher frequency, to the point that the true distribution of the frequency bins cannot be adequately resolved. The Galactic confusion for the simulation set is computed by averaging over all noise side realizations in the set.

\begin{figure}[htb]
\includegraphics[width=\columnwidth]{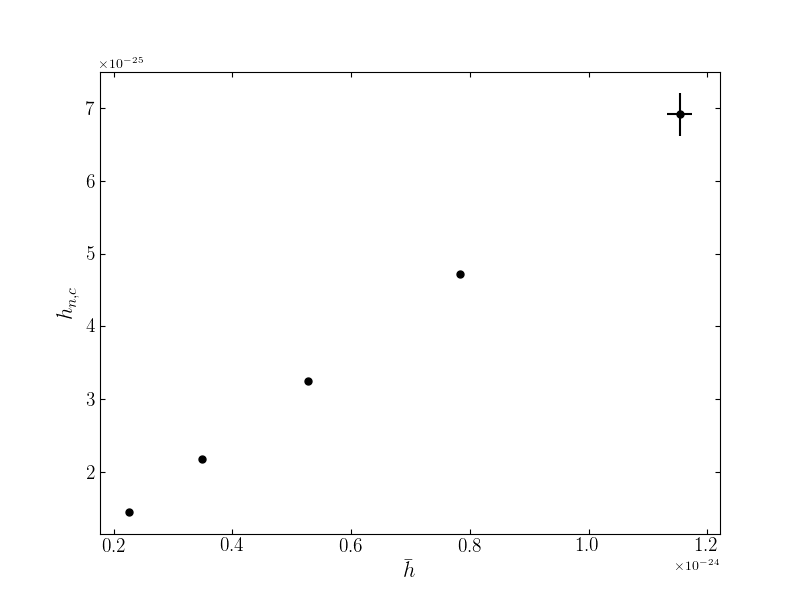}
\caption{A plot of $h_{\mathrm{n,c}}$ against $\bar{h}$ for $G\mu=10^{-18}$ and $\alpha=0.1$, at $f=0.05\ \mathrm{Hz}$, computed for the simulation sets including the closest $\left[10^1, 10^2, 10^3, 10^4, 10^5\right]$ loops, averaged over $[96, 96, 96, 96, 10]$ realizations in each set, respectively.}
\label{fig:clmjust}
\end{figure}

We directly test the closest loops method by computing $h_{\mathrm{n,c}}$ and $\bar{h}$ for $G\mu=10^{-18}$ and $\alpha=0.1$, for the simulation sets including the closest $\left[10^1, 10^2, 10^3, 10^4, 10^5\right]$ loops, averaged over $[96, 96, 96, 96, 10]$ realizations in each set, respectively. We present an example at $f=0.05\ \mathrm{Hz}$ in fig. \ref{fig:clmjust}, because this is in the frequency range where the presence of signals is common. The 5 data points from lower to higher strain amplitudes corresponding to the growing number of loops included display a good linear relationship as required by eq. \ref{eqn:clm}. By utilizing the closest loops method, we essentially extrapolate the line at each frequency to compute $h_{\mathrm{n,c}}$ for the entire Galaxy. This linear relationship holds down to a subset of just 10 loops and does start breaking down for even smaller subsets. Thus, our noise side simulation sets including 1000 loops are indeed sufficient for recovering the full Galactic $h_{\mathrm{n,c}}$.

\begin{figure}[htb]
\includegraphics[width=\columnwidth]{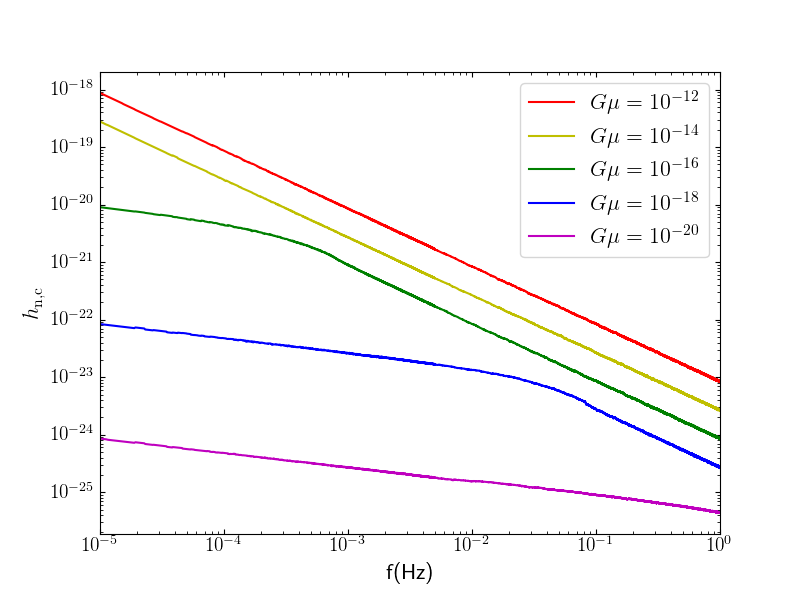}
\caption{The full $h_{\mathrm{n,c}}$ of the Galaxy with the rocket effect and $\alpha=0.1$, computed using the closest loops method.}
\label{fig:hnctot}
\end{figure}

Note that the cliff in fig. \ref{fig:hncexp} is in a sense an artificial feature caused by including only a subset of loops. It therefore must go away if we correctly recover the full Galactic $h_{\mathrm{n,c}}$ through the closest loops method. This is indeed the case in fig. \ref{fig:hnctot} showing the full $h_{\mathrm{n,c}}$ of the Galaxy with the rocket effect and $\alpha=0.1$ computed using the closest loops method. Clearly, the small breaks at $f_{\odot}$ are fully recovered, and the shapes of the curves resemble those in fig. \ref{fig:galhtho} as expected.

\subsubsection{The Stochastic Background}
\label{sssec:sto}
The stochastic background of GW from loops can be computed by redshifting and integrating the power of GW emitted by all loops over all times, from the formation time of the earliest loops whose GW becomes relevant to the LISA frequency range, $t_i$\cite{DP07},
\begin{equation}
\label{eqn:storho}
\rho(f)=\Gamma G\mu^2\int_{t_i}^{t_0}dt\left(\frac{a(t)}{a(t_0)}\right)^4n\left(f\frac{a(t_0)}{a(t)},t\right).
\end{equation}
As a background energy density, it is customarily expressed\cite{DP07,Moore2015} as an energy density spectrum in terms of the critical density $\rho_{\mathrm{c}}=\frac{3H_0^2}{8\pi G}$,
\begin{equation}
\label{eqn:stoomega}
\Omega_{\mathrm{g}}(f)=\frac{8\pi G}{3H_0^2}\frac{d\rho(f)}{d\ln f}.
\end{equation}
The fact that eq. \ref{eqn:storho} concentrates the total power of emission in the fundamental mode is unimportant for our purpose, as the stochastic background is not significantly changed by GW emission spectra\cite{DP07}. Higher frequency modes are incorporated when necessary.

Among the various quantities commonly used for expressing GW, the characteristic strain $h_{\mathrm{c}}$ is usually the most convenient to work with for data analysis, as it automatically incorporates the integration time over which the signal is observed. Eq. \ref{eqn:stoomega} can be converted using\cite{Moore2015}
\begin{equation}
h_{\mathrm{c,u}}(f)=\frac{H_0}{\pi f}\sqrt{\frac{3\Omega_{\mathrm{g}}(f)}{2}}.
\end{equation}
Since the stochastic background is really an unresolved GW background energy density, its actual strain amplitude cannot be simply defined, nor is that necessary as we always work with $h_{\mathrm{c}}$ in our data analysis. However, as we will see in section \ref{sssec:SNRcalc}, it is convenient practically to define an effective strain amplitude for the stochastic background
\begin{equation}
\label{eqn:hnusto}
h_{\mathrm{n,u}}(f)\equiv\frac{h_{\mathrm{c,u}}(f)}{\sqrt{T_{\mathrm{obs}}f}}.
\end{equation}

\begin{figure}[htb]
\includegraphics[width=\columnwidth]{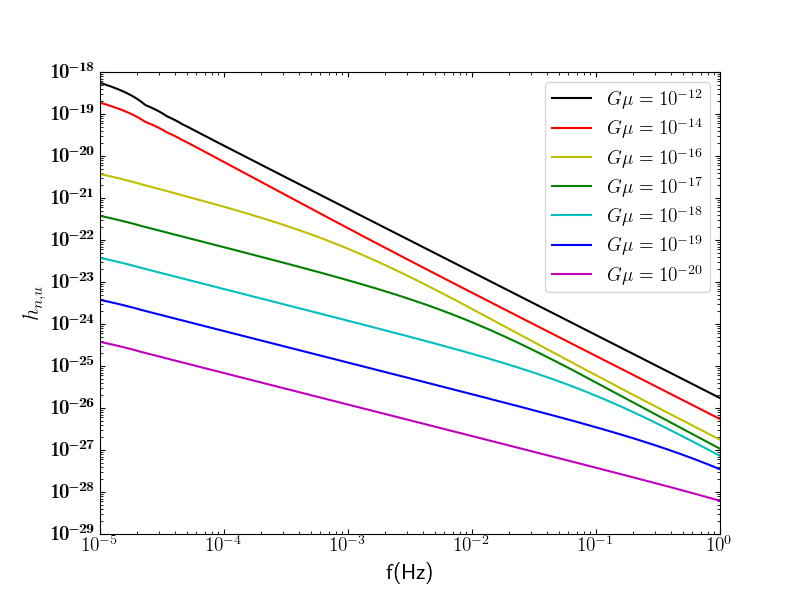}
\caption{the stochastic background plotted as $h_{\mathrm{n,u}}$ with $\alpha=0.1$ and $T_{\mathrm{obs}}=1\ \mathrm{year}$.}
\label{fig:hnu}
\end{figure}

We plot $h_{\mathrm{n,u}}$ in fig. \ref{fig:hnu} with $\alpha=0.1$ and $T_{\mathrm{obs}}=1\ \mathrm{year}$. The change in slope indicates the transition into the high frequency flat region when plotted as $\Omega_{\mathrm{g}}(f)$, where loop decay is significant.

\subsubsection{The Instrumental Noise and the Galactic WD Binaries Background}
\label{sssec:hLISA}
\begin{figure}[htb]
\includegraphics[width=\columnwidth]{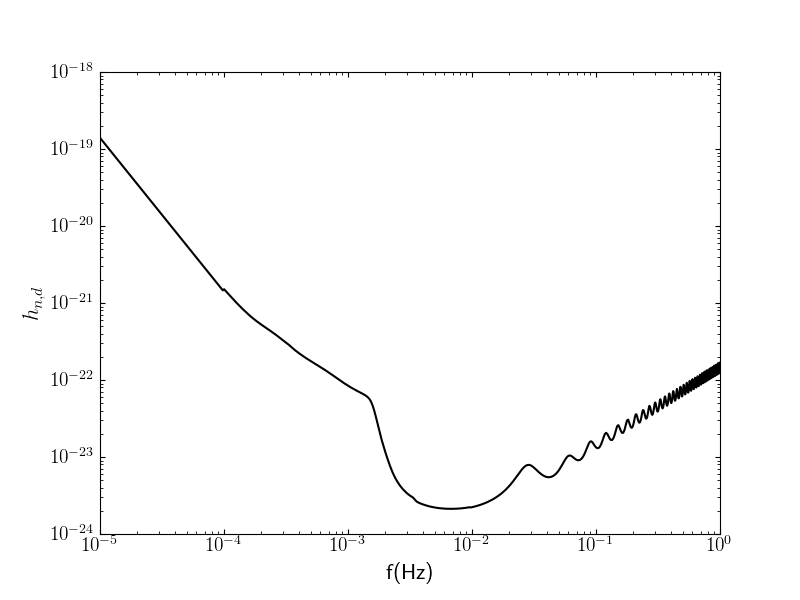}
\caption{The all-sky LISA instrumental noise with the standard detector configuration averaged over polarizations\cite{LISAsens} merged with the Galactic WD binaries background\cite{WD}, plotted as $h_{\mathrm{n,d}}$ with $T_{\mathrm{obs}}=1\ \mathrm{year}$.}
\label{fig:hLISA}
\end{figure}

As a space-based GW detector, the orbiters of LISA experience considerable variations in their relative positions capable of mimicking GW signal\cite{LISAwp}, setting the floor of LISA sensitivity. The level of signal necessary so that a signal-to-noise ratio (SNR) of unity can be obtained from the LISA observation is defined to be its sensitivity curve, which is then interpreted as a source of noise, i.e. the instrumental noise, for data analysis. We adopt the all-sky LISA instrumental noise\cite{LISAsens} averaged over polarizations with a standard triangular detector configuration. This LISA instrumental noise accounts for most of the features in fig. \ref{fig:hLISA} which is plotted as the effective strain amplitude $h_{\mathrm{n,d}}$ with $T_{\mathrm{obs}}=1\ \mathrm{year}$, including the overall trend and the ringing at high frequency. The same as in section \ref{sssec:sto}, $h_{\mathrm{n,d}}$ is defined in terms of $h_{\mathrm{c,d}}$ as
\begin{equation}
\label{eqn:hndLISA}
h_{\mathrm{n,d}}(f)\equiv\frac{h_{\mathrm{c,d}}(f)}{\sqrt{T_{\mathrm{obs}}f}}.
\end{equation}

A major astrophysical source of noise in the LISA frequency range consists of the WD binaries population which is extremely abundant in the Galaxy, with an abundance in the halo estimated to be\cite{WD} $N_{\mathrm{WD}}\sim 10^{12}$ in the LISA frequency range. They are therefore an important source of GW in the Galaxy while remaining mostly unobservable otherwise. Far from the end of their inspirals, WD binaries in this frequency range are essentially monochromatic GW emitters, making them particularly relevant for our purpose. Their GW forms an unresolved confusion background primarily contributing to $f\lesssim 10^{-3}\ \mathrm{Hz}$ as can be seen in fig. \ref{fig:hLISA}. The fact that Galactic WD binaries mostly affect the low frequency region diminishes their relevance for our purpose, because of the relatively low loop abundance in that region.

\subsubsection{The Signal-to-Noise Ratio}
\label{sssec:SNRcalc}
The general optimal SNR for data analysis of GW detection is\cite{SNR}
\begin{equation}
\label{eqn:snropt}
\varrho^2=4\int_0^{\infty}df\frac{\left|\tilde{h}(f)\right|^2}{S_{\mathrm{n}}(f)},
\end{equation}
where $\tilde{h}(f)$ is the Fourier transform of the time-domain signal strain, and $S_{\mathrm{n}}(f)$ is the noise root spectral density. For monochromatic sources, the Dirac delta function in $\tilde{h}(f)$ kills the integral, and the SNR greatly simplifies to\cite{SNR}
\begin{equation}
\label{eqn:snrmono}
\varrho^2(f)=\frac{h(f)^2T_{\mathrm{obs}}}{S_{\mathrm{n}}(f)}.
\end{equation}

To proceed further, we turn our attention to $h_{\mathrm{c}}$ of monochromatic sources. Binary inspirals are quasi-monochromatic sources of GW with\cite{Moore2015, inspiral}
\begin{equation}
\label{eqn:ib5}
h_{\mathrm{c}}(f)=h(f)\sqrt{N_{\mathrm{cyc}}},
\end{equation}
where $N_{\mathrm{cyc}}$ is the number of wave cycles the binary inspiral emits GW in the frequency bin $f$, reflecting the limit on signal integration. For monochromatic sources, signal integration is instead limited by $T_{\mathrm{obs}}$, and therefore the correct identification of $h_{\mathrm{c}}$ is then\cite{Moore2015}
\begin{equation}
\label{eqn:hcmono}
h_{\mathrm{c}}(f)=h(f)\sqrt{N_{\mathrm{obs}}}=h(f)\sqrt{T_{\mathrm{obs}}f}.
\end{equation}

Now the SNR for monochromatic sources can be derived by substituting eq. \ref{eqn:hcmono} into eq. \ref{eqn:snrmono} and converting $S_{\mathrm{n}}$ into $h_{\mathrm{c,n}}$,
\begin{equation}
\label{eqn:snrchar}
\varrho^2(f)=\frac{h(f)^2T_{\mathrm{obs}}f}{h_{\mathrm{c,n}}(f)^2}=\left(\frac{h_{\mathrm{c,s}}(f)}{h_{\mathrm{c,n}}(f)}\right)^2.
\end{equation}
The SNR for monochromatic sources is simply given by the ratio of the signal and the noise $h_{\mathrm{c}}$. Dividing both the numerator and the denominator by $N_{\mathrm{obs}}$, schematically, the SNR in our simulations is computed as
\begin{equation}
\label{eqn:snrreal}
\varrho=\frac{h(f)}{\sqrt{\sum_i h_{\mathrm{n,i}}(f)^2}},
\end{equation}
where the sum is over all sources of noise considered. As loops emitting harmonic GW are persistent sources in the Galaxy, we require a relatively low
\begin{equation}
\label{eqn:snrreq}
\varrho\geq 3
\end{equation}
for an event to be counted.

\subsection{Results}
\label{ssec:toyresults}
\subsubsection{Signal from Nearby Loops}
\label{sssec:decetcomp}
\begin{figure}[htb]
\includegraphics[width=\columnwidth]{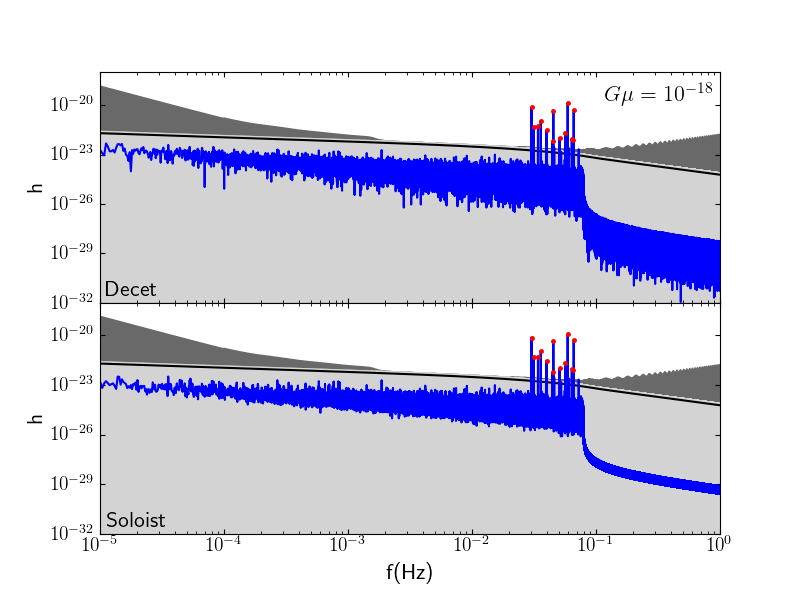}
\caption{Snapshots of a realization in the signal side simulation set with the rocket effect at $G\mu=10^{-18}$ and $\alpha=0.1$. The upper and lower panels display signals for the same realization including the closest 10 loops (the ``decet'') and the single loop (the ``soloist'') for each frequency bin, respectively. For both panels, the black curve is $\bar{h}$, and the shaded regions from lighter to darker shades of grey represent the cumulative inclusion of sources of noise from: the Galactic confusion noise, the stochastic background, and the LISA instrumental noise with the Galactic WD binaries background, respectively. Red dots indicate events satisfying the SNR requirement.}
\label{fig:decetcomp}
\end{figure}

We present here results illustrating the validity of the closest loops method on the signal side. Snapshots from a realization in the signal side simulation set with the rocket effect at $G\mu=10^{-18}$ and $\alpha=0.1$ are plotted in fig. \ref{fig:decetcomp}. For visual clarity, we select the realization with the lowest number of events satisfying the SNR requirement for this purpose. The total power of GW emission is concentrated in the fundamental mode to avoid unnecessary complications irrelevant to the central issue of this section. The upper panel reflects the normal method with which our simulation is performed and its results analyzed, where the closest 10 loops for each frequency bin are included in the signal side. There are 13 events in total. The data analysis procedure is repeated in the lower panel for the same realization, with the signal for each frequency bin coming from only the closest single loop. There are again 13 events in total, as expected.

\begin{figure}[htb]
\includegraphics[width=\columnwidth]{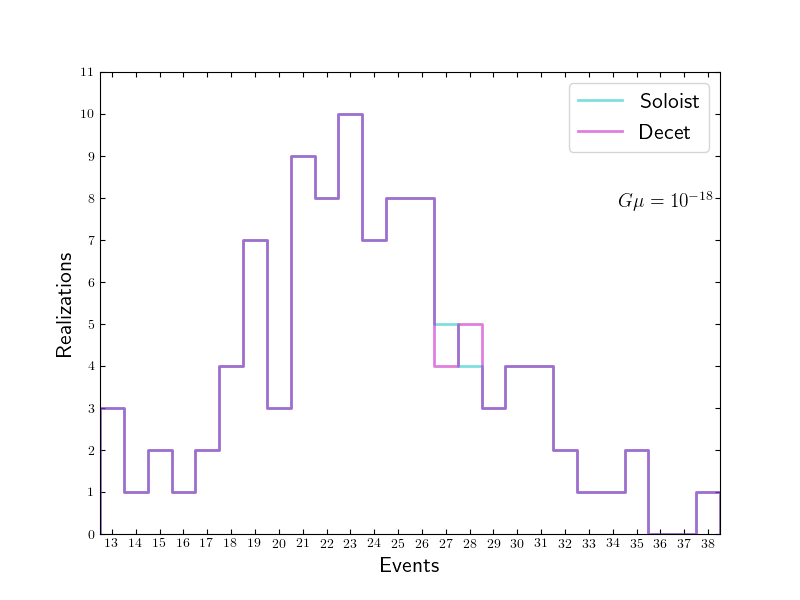}
\caption{Histograms of events satisfying the SNR requirement for the simulation set with the rocket effect at $G\mu=10^{-18}$ and $\alpha=0.1$. The histograms in magenta and cyan include the closest 10 loops and the single loop for each frequency bin, respectively.}
\label{fig:decethisto}
\end{figure}

A more conclusive validation can be obtained by comparing statistically results for all realizations in the simulation set. The histograms in fig. \ref{fig:decethisto} cover the same simulation set with the rocket effect at $G\mu=10^{-18}$ and $\alpha=0.1$, with the difference that one (magenta) is generated by including contribution from the closest 10 loops for each frequency bin for the signal side, while the other (cyan) only includes signal from the closest single loop. Visually, it is clear that they are essentially identical, with the exception of a single event in a single realization which appears to be absent in the latter case. Statistically, the numbers of events from both histograms are also identical,
\begin{eqnarray}
\label{eqn:events1m}
E_{\mathrm{D}} & = & 24\pm 5; \\
E_{\mathrm{S}} & = & 24\pm 5.
\end{eqnarray}
This comparison serves as a clear illustration that most events are in fact due to just a single loop being located extremely closely, and not from the combined contribution of GW emitted by multiple nearby loops.

The simulation set chosen is in the relevant region of the parameter space, where loop abundance in the Galaxy is very high, and the number of events is plentiful, ensuring the relevance and generality of the result obtained. Physically, the result means that even with such high loop abundance in the toy model, the probability that a single loop happens to be located sufficiently nearby to be resolved is still quite low, to the point that one has to be very lucky to have just one such occurrence. As the probability of multiple such occurrences in the same frequency bin decreases exponentially, tracking the closest 10 loops for each frequency bin for the signal side safely ensures that our simulation accounts for all potential event candidates in the Galaxy. Coupled with the noise side validation from section \ref{sssec:hconf}, this result makes sure that despite constraints on computational capabilities, our simulation is a reasonable representation of loops in the Galaxy for the purpose of this study.

\subsubsection{Gravitational Wave Emission Spectrum}
\label{sssec:GWspec}
The computationally viable means of incorporating higher harmonic modes of GW emission from loops into our simulation is through postprocessing. The procedure is implemented from lower to higher frequencies, and consists of first renormalizing the fundamental strain amplitude according to eq. \ref{eqn:Pnnorm}, then distributing the power into higher harmonics according to eqs. \ref{eqn:cuspspec} and \ref{eqn:kinkspec}, and lastly computing the superposition of these higher harmonics with existing GW at those frequencies. This procedure avoids the burden of tracking higher harmonics for each loop, enabling us to incorporate different GW emission spectra easily and quickly.

A disadvantage of the above method is that the very low frequency region is missing contribution from higher harmonics of GW emitted by loops whose fundamental modes are below the frequency range, with the primary effect being a slight reduction in the Galactic confusion in this region. This is unimportant for our purpose for several reasons. First, the reduction is not very significant because the fundamental mode dominates in any GW emission spectrum, and after distributing power to higher harmonics, the fundamental only $h$ itself is only reduced by a factor of $\sim\sqrt{2}$. Second, $h_n$ decreases rapidly as $n$ becomes large, making the underestimation to the Galactic confusion very slight, as shown in ref. \cite{DP09}. Third, this rapid decrease of $h_n$ means that any effect is only relevant for a narrow band in the low frequency range $1\times 10^{-5}\ \mathrm{Hz}\leq f\lesssim 5\times 10^{-5}\ \mathrm{Hz}$. As will become clear in section \ref{sssec:rocket}, this low frequency band is irrelevant for the simulation sets in the important part of the parameter space, due to a combination of low loop abundance and the fact that in these sets, the instrumental noise with the Galactic WD binaries background greatly dominates over the Galactic confusion at those frequencies. One final argument for the unimportance of the missing contribution at very low frequency is the fact that the effects of higher harmonics themselves are actually minimal for the purpose of our simulation, as will become clear below.

\begin{figure}[htb]
\includegraphics[width=\columnwidth]{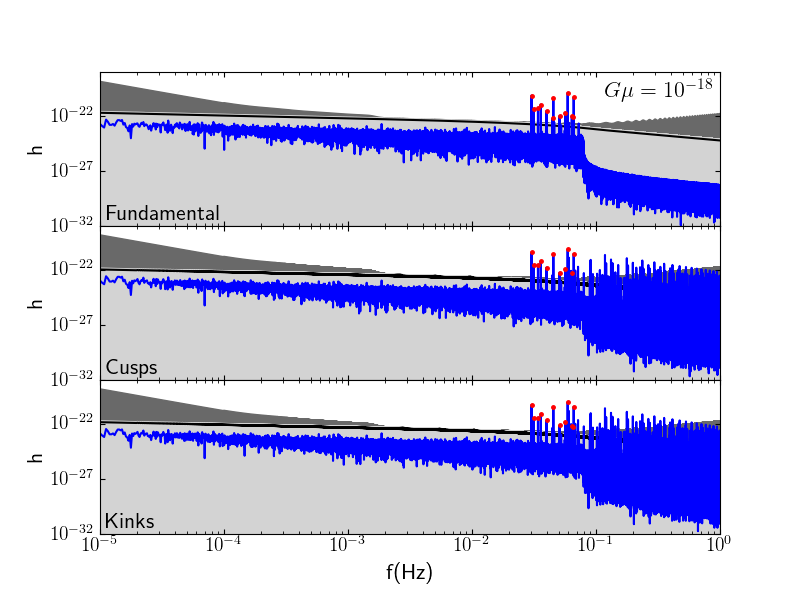}
\caption{Snapshots of the same realization as that shown in fig. \ref{fig:decetcomp}. Please refer to the caption there for an explanation of the layout of the plot. The panels from top to bottom display snapshots of this realization with three distributions of the power of GW emission by loops: all concentrated in the fundamental mode, distributed according to the cusps-dominated spectrum, distributed according to the kinks-dominated spectrum, respectively.}
\label{fig:speccomp}
\end{figure}

An example illustrating any effects GW emission spectra may have on our simulation is presented in fig. \ref{fig:speccomp}, showing snapshots of the same realization selected in fig. \ref{fig:decetcomp}. The top panel displays the signal when all the power of GW emission is concentrated in the fundamental mode, and is identical to the upper panel of fig. \ref{fig:decetcomp}, with 13 events in total. The lower two panels display snapshots of the same realization when some power of GW emission is distributed in higher harmonics, with loops in the middle and bottom panels having the cusps- and the kinks- dominated spectra, respectively. We do not require coincident detection of higher harmonics of a fundamental mode signal here for a fair comparison. Higher harmonics of events whose fundamental modes have already been detected are carefully avoided, which are clearly visible at high frequency in the lower two panels. There are 11 and 12 events in total for signals with the cusps- and the kinks-dominated spectra, respectively. The slight differences in the total numbers of events for the three panels can be understood quite simply. The SNR is of course maximized when all the power is concentrated in the fundamental mode, and hence the highest number of events in that case. The SNR is reduced when some power is distributed in higher harmonics, and the effect is more pronounced for the shallower cusps-dominated spectrum, giving the lowest number of events.

\begin{figure}[htb]
\includegraphics[width=\columnwidth]{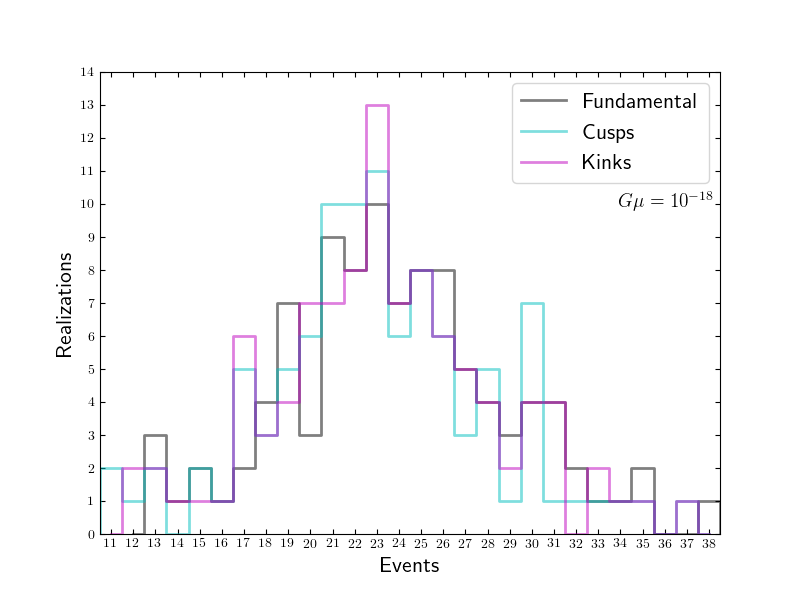}
\caption{Histograms of events for the same simulation set as that shown in fig. \ref{fig:decethisto}. The histogram in grey is for when all the power of GW emission is concentrated in the fundamental mode. Those in cyan and magenta are for when some power is distributed in higher harmonics according to the cusps- and the kinks-dominated GW emission spectra, respectively.}
\label{fig:spechisto}
\end{figure}

A full comparison of the three cases for all realizations in the simulation set can be made statistically, with histograms shown in fig. \ref{fig:spechisto} for the same set as that shown in fig. \ref{fig:decethisto}. Minor differences in the detailed features notwithstanding, the histograms for the three GW emission spectra have similar overall shapes and statistical properties, with the total numbers of events
\begin{eqnarray}
E_{\mathrm{F}} & = & 24\pm 5; \\
E_{\mathrm{C}} & = & 23\pm 5; \\
E_{\mathrm{K}} & = & 23\pm 5.
\end{eqnarray}
Unsurprisingly, $E_{\mathrm{F}}$ is the highest as the SNR is maximized there. But the important point here is that GW emission spectra do not have significant effects on our simulation, which statistically is not at all sensitive to them when no coincident detection is required.

\subsubsection{Coincident Detection of Harmonic Modes}
\begin{figure}[htb]
\includegraphics[width=\columnwidth]{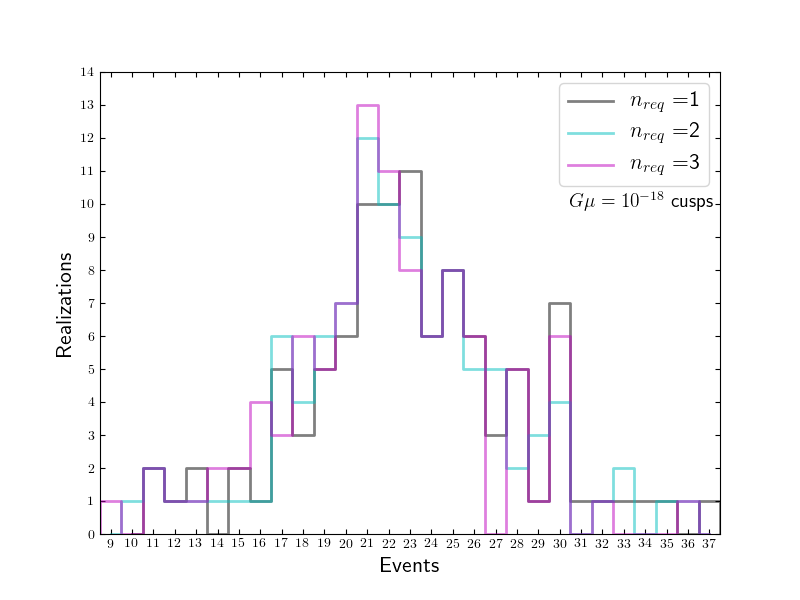}
\caption{Histograms of events for the same simulation set as that shown in fig. \ref{fig:decethisto}, with the cusps-dominated GW emission spectrum. The histogram in grey is for when coincident detection of higher harmonics is not required. Those in cyan and magenta are for when coincident detections of the the first one and two harmonics are required, respectively.}
\label{fig:cuspshisto}
\end{figure}

\begin{figure}[htb]
\includegraphics[width=\columnwidth]{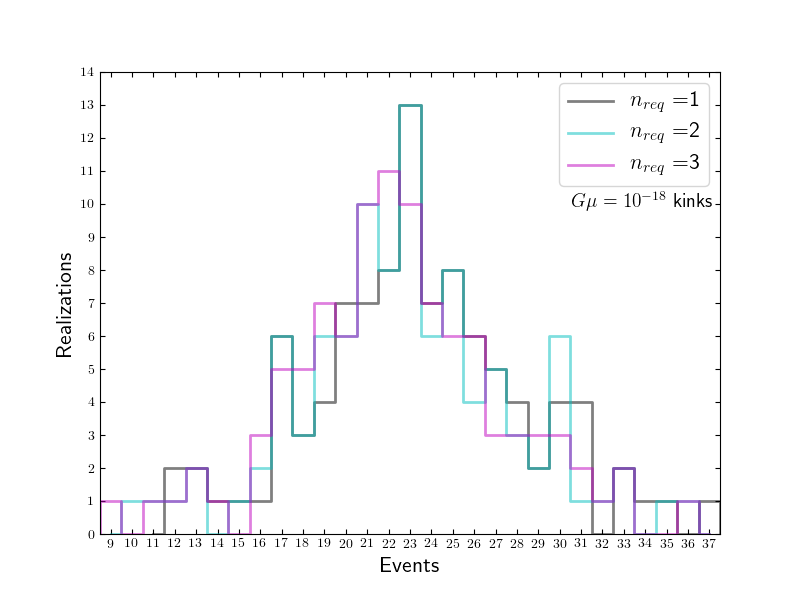}
\caption{The histograms are the same as those in fig. \ref{fig:cuspshisto} except that GW emission is kinks-dominated.}
\label{fig:kinkshisto}
\end{figure}

Since the harmonic signature is unique to GW from loops not replicated by any other sources, it is important to take advantage of it in the search of harmonic signal from resolved loops. As discussed in section \ref{sssec:hLISA}, WD binaries in the Galaxy are the most capable of causing confusion with loops. While the confusion background formed by these sources have been incorporated, with an abundance in the Galaxy as high as $N_{\mathrm{WD}}\sim 10^{12}$ in the LISA frequency range\cite{WD}, they possibly have the potential to mimic the signal as well, should one happen to be located extremely nearby. While a detailed study of the possibility of detecting such signal from WD binaries would be beneficial for its own merit, it is not necessary for our purpose, because WD binaries are not harmonic sources. While they are also persistent sources of essentially monochromatic GW, they are unable to consistently mimic the harmonic signature of loops, and therefore can easily be excluded by requiring coincident detection of higher harmonics for candidate events. Even though $N_{\mathrm{WD}}$ appears high, our results (fig. \ref{fig:decetcomp}) have already shown that even with a total number of loops in the Galaxy that is orders of magnitude higher in the toy model, the probability that a loop can be resolved out of even just the confusion of other loops in the Galaxy is still very low. This means that we would be lucky to individually resolve a few Galactic WD binaries, and the likelihood of multiple such sources consistently mimicking the harmonic signature is negligible.

While requiring coincident detection of higher harmonics can effectively eliminate false events caused by potentially confusing signal, such as that from Galactic WD binaries, genuine events should not be significantly affected as long as we do not require coincident detection of an excessive number of harmonic modes. We verify this by comparing statistical results for the same simulation set as that shown in fig. \ref{fig:decethisto}, with different coincident detection requirements. The histograms with the cusps- and the kinks-dominated GW emission spectra are presented in figs. \ref{fig:cuspshisto} and \ref{fig:kinkshisto}, respectively, with the requirements for coincident detection set as no requirement (grey), at most the 1st harmonic mode (cyan), at most the 2nd harmonic mode (magenta). Histograms of the first case correspond to cyan and magenta histograms in fig. \ref{fig:spechisto}. The SNR requirement is modified according to the relative strengths of higher harmonics.

Minor differences in the detailed features notwithstanding, histograms in each figure all have similar shapes and statistical properties. It is somewhat noticeable that magenta histograms appear to have peaks shifted slightly to the left relative to the others. They give us the following event numbers
\begin{eqnarray}
E_{\mathrm{C,1}} & = & E_{\mathrm{K,1}}=23\pm 5; \\
E_{\mathrm{C,2}} & = & E_{\mathrm{K,2}}=23\pm 5; \\
E_{\mathrm{C,3}} & = & E_{\mathrm{K,3}}=22\pm 5.
\end{eqnarray}
The results show that requiring coincident detection of the first harmonic appears to be the sweet spot, where we can reap the benefit of being able to eliminate false events without significantly affecting legitimate ones. We therefore adopt this requirement for the toy model unless otherwise specified, with the cusps-dominated GW emission spectrum, as cusps are generally dominant.

\subsubsection{String Tension}
\label{sssec:tension}
To analyze the effects varying $G\mu$ has on the results, we compare statistical results of event numbers obtained from histograms similar in nature to those presented earlier, for the simulation sets at $\alpha=0.1$ without the rocket effect and with $G\mu$ varying from $10^{-12}$ to $10^{-20}$. They are summarized in the second column in table \ref{tab:events}. As will become clear in sections \ref{sssec:rocket} and \ref{sssec:alpha}, this selection of simulation sets maximizes the total numbers of events, thereby giving the most informative results, while precluding complications not pertinent to the main issue discussed presently.

The most prominent feature of these statistical results is the relatively large number of events for $G\mu=10^{-12}$. Then the number of events just falls off a cliff as $G\mu$ decreases, before it starts increasing again and peaks at $G\mu=10^{-18}$, after which it again drops. To explain this feature, we note that a large number of events can be generated in two ways. The first way is to have strong GW emission from each loop. This effectively makes sources of noise other than the Galactic confusion less relevant, and the harmonic signal from a nearby loop can be detected without requiring it to be located in extreme proximity, as the signal only has to stand out over the Galactic confusion. Since the probability that a loop can overwhelm the Galactic confusion is not dependent on the strength of GW emission from each loop, and that $h\propto G\mu$ from eq. \ref{eqn:hindloop}, this way favors higher $G\mu$ to ensure that other sources of noise are subdominant. The second way is to have high loop abundance in the Galaxy, which increases the likelihood that a loop happens to be present in extreme proximity that it overwhelms all sources of noise. From fig. \ref{fig:Nnr}, this way favors lower $G\mu$ up to a point. These two ways mean that the event number is a result of the tension between the strength of GW emission of each loop and loop abundance in the Galaxy, and that the two sides favor higher and lower $G\mu$, respectively. The ideal scenario for detection would of course be to have both. But fig. \ref{fig:Nnr} tells us that we do not have that case, not even in the toy model. This tension explains features observed when $G\mu$ is varied.

At $G\mu=10^{-12}$, the Galactic confusion dominates the noise, and GW emission from each loop is the strongest. Thus, despite low loop abundance in the Galaxy, detecting harmonic signal from nearby loops is not very difficult as they just have to be nearby relative to the rest of the loop population at their frequency, and hence the large event number. As $G\mu$ decreases, the strength of GW emitted by each loop decreases, and other sources of noise take over. It now becomes impossible to detect harmonic signal from nearby loops that are just relatively nearby, while loop abundance is still far too low to enable an appreciable number of loops to be located in extreme proximity, and hence the very low event numbers at $G\mu=10^{-14}$ and $10^{-16}$. Loop abundance increases as $G\mu$ decreases further, contributing to a higher number of events and reaching a peak at $G\mu=10^{-18}$. Such loops with very low $G\mu$ decay slowly, with those in the LISA frequency range not significantly affected by loop decay. The resultant great loop abundance enables many to be located extremely closely, to the point that their harmonic signal becomes detectable despite the relatively weak GW emission. The number of events drops again as $G\mu$ decreases even further, due to less significant enhancement to loop abundance. At a given frequency, loop abundance does not depend on $G\mu$ intrinsically as can be seen from eq. \ref{eqn:ininumdens}, and stops increasing as $G\mu$ decreases once loop decay ceases being significant. This means that while the requirement of proximity for detection gets higher at $G\mu=10^{-19}$ and $10^{-20}$ due to weaker GW emission, these loops are not much more likely to be located more closely, lowering the event numbers. More importantly, the event number will only become even lower should $G\mu$ decrease more.

These results mean that when the rocket effect is not considered, $G\mu\lesssim 10^{-17}$ appears to be the best region in the parameter space, while $G\mu\sim 10^{-12}$ also seems viable.

\subsubsection{The Rocket Effect}
\label{sssec:rocket}
As the rocket effect ejects all loops in the Galaxy beyond $r_{\mathrm{tr}}$, our expectation for its effects on simulation results is simply that no detection should be possible when $f>f_{\odot}$, because no loop can be located particularly closely as the entire population is bound by $r_{\mathrm{tr}}$. This strict elimination of signal should severely constrain detection for simulation sets with all but the lowest $G\mu$.

We verify this by analyzing statistical results of the simulation sets with the rocket effect, again at $\alpha=0.1$ so that they can be compared with results presented earlier, with $G\mu$ also varying from $10^{-12}$ to $10^{-20}$. They are summarized in the first column in table \ref{tab:events}.

As expected, all events are eliminated by the rocket effect at $G\mu=10^{-12}$. This is especially significant given that the event number is high without the rocket effect, and is a direct result of the fact that here $f>f_{\odot}\forall f$ in the frequency range. For sets with lower $G\mu$, $f_{\odot}$ increases, and the impact of the rocket effect diminishes, as larger portions of the frequency range become viable for detection. We therefore start recovering significant numbers of events with $G\mu\lesssim 10^{-18}$. Recall from section \ref{sssec:rockettho}, that $f_{\odot}$ tracks the beginning of the decay regime as frequency increases. Thus, just as is the case for loop decay, the rocket effect also starts becoming irrelevant for $G\mu\lesssim 10^{-19}$. For this reason, event numbers at $G\mu\lesssim 10^{-19}$ are similar with and without the rocket effect. The small enhancement at $G\mu=10^{-20}$ with the rocket effect may be attributed to the fact that when $r_{\mathrm{tr}}\gtrsim r_{\odot}$, the rocket effect is actually beneficial to detection due to the reduction in the Galactic confusion.

The rocket effect eliminates $G\mu=10^{-12}$ as a viable region in the parameter space, and the sweet spot for detection is therefore $G\mu\lesssim 10^{-18}$, which should be the focus for physical results.

\subsubsection{Initial Loop Sizes}
\label{sssec:alpha}
\begin{figure}[htb]
\includegraphics[width=\columnwidth]{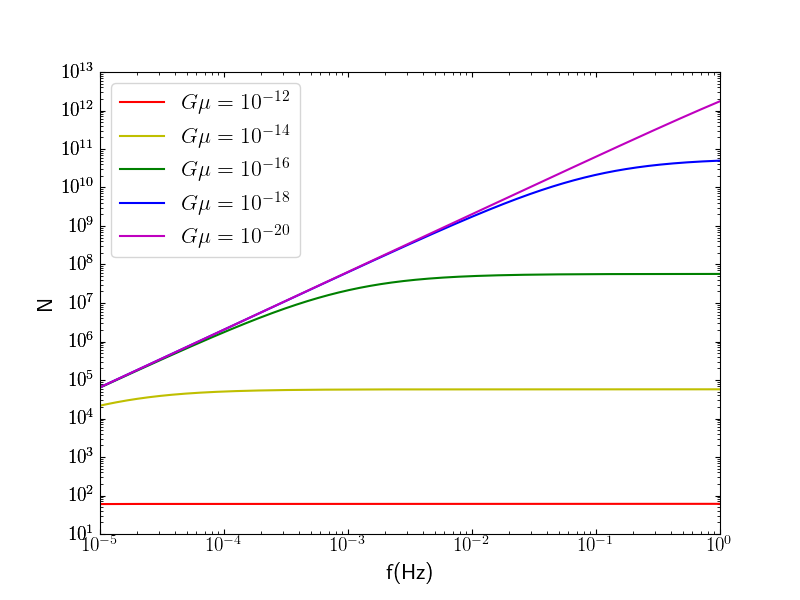}
\caption{The same as fig. \ref{fig:Nnr} except that $\alpha=10^{-5}$.}
\label{fig:sNnr}
\end{figure}

Eq. \ref{eqn:ininumdens} appears to suggest that loop abundance should increase with a smaller $\alpha$, which is true for loops with a given $t_{\mathrm{c}}$. But when the frequency range is fixed, decreasing $\alpha$ increases $t_{\mathrm{c}}$, and actually reduces loop abundance because of the larger $H\left(t_{\mathrm{c}}\right)^{-3}$. This fact is clearly reflected by comparing loop abundance with $\alpha=10^{-5}$ shown in fig. \ref{fig:sNnr} to that with $\alpha=0.1$ in fig. \ref{fig:Nnr}. The former is consistently lower than the latter by about two orders of magnitude. We again exclude the rocket effect here to eliminate complications from effects not pertinent to the issue discussed presently. Another feature to note from the comparison is the fact that the decay regimes remain the same, because they are really determined by the relative values of $\alpha t_{\mathrm{c}}$ and $\Gamma G\mu\left(t_0-t_{\mathrm{c}}\right)$. Recall from section \ref{sssec:rockettho}, the first quantity is independent of $\alpha$. For the second quantity, even though overall loops with $\alpha=10^{-5}$ are younger, the difference in $t_{\mathrm{c}}$ is negligible when compared to $t_0$. For the same reason, $f_{\odot}$ also remains similar and tracks the decay regime. The primary difference between the simulation sets with different $\alpha$ for our purpose manifests in differences in loop abundance in the Galaxy.

We now ask the question, for a given frequency bin, how does changing loop abundance in the Galaxy affect the probability of detection? A detailed answer is likely quite complicated, and is not essential to our study. Heuristically, there are three scenarios depending on the relative dominance of the Galactic confusion among all sources of noise. The first scenario applies when other sources of noise strongly dominate, and detections are very low probabilistic events, with any contribution being essentially due to the shell of closest loops, which has $N^{\frac{2}{3}}$. The probability of detection should be proportional to the mean scattering of the distribution of distances to these loops, which is a random walk with the mean proportional to $\sqrt{N^{\frac{2}{3}}}=N^{\frac{1}{3}}$. Since the overall noise level remains fixed as $N$ varies, we need to take into account the contribution from geometry, and the probability should also be inversely proportional to the average distance to these loops, i.e.
\begin{equation}
\label{eqn:Pbin1}
p_{k}\propto\frac{s_{\mathrm{shell}}}{d_{\mathrm{shell}}}\propto\frac{N^{\frac{1}{3}}}{N^{-\frac{1}{3}}}=N^{\frac{2}{3}},
\end{equation}
where $k$ is the frequency bin. The second scenario applies when other sources of noise only weakly dominate over or are comparable to the Galactic confusion. It is now easier for nearby loops to be detected, and instead of a shell, contribution comes from the volume of nearby loops. Thus, the scattering in the distribution of distances now scales as $\sqrt{N}$, and the probability of detection becomes
\begin{equation}
\label{eqn:Pbin2}
p_{k}\propto\frac{s_{\mathrm{vol}}}{d_{\mathrm{vol}}}\propto\frac{N^{\frac{1}{2}}}{N^{-\frac{1}{3}}}=N^{\frac{5}{6}}.
\end{equation}
The third scenario applies when the Galactic confusion dominates. Detections are relatively easy, with contribution still coming from the nearby volume with scaling $\sqrt{N}$. Unlike the other two scenarios however, the overall noise now changes with $N$. Combining eqs. \ref{eqn:hbarN} and \ref{eqn:hncN}, the noise also scales as $\sqrt{N}$ modulated by geometry. We therefore disregard geometry, and see that we do not expect $p_{k}$ to depend strongly on loop abundance in this case.

We analyze effects of changing $\alpha$ on the simulation by computing statistical results for the simulation sets without the rocket effect and with $\alpha=10^{-5}$, summarized in the fourth column in table \ref{tab:events}.

Though the simulation sets with $\alpha=10^{-5}$ have more events than those with $\alpha=0.1$ when $G\mu$ is high, by a factor of $\sim 3$, they are still consistent with the third scenario of the heuristic argument above, given that loop abundance differs by about two orders of magnitude. Moreover, as these sets contain relatively few loops, with only $N\sim \mathcal{O}(60)$ at each frequency when $G\mu=10^{-12}$, the Galactic confusion itself becomes less well-defined. In this case, conceivably it is relatively easy for a loop to become detectable by virtue of being located just somewhat more closely relative to the others, without actually being very close, thereby generating some minor enhancement in the event number.

Just as with results with $\alpha=0.1$, as $G\mu$ decreases, the event number first drops and then stages a rebound. However, here the rebound is far less significant in comparison, which comes as a direct result of the lower loop abundance. Thus, the only potentially viable region in the parameter space with $\alpha=10^{-5}$ is $G\mu\sim 10^{-12}$. We can already sense that when coupled with the rocket effect, this result implies that having smaller $\alpha$ is disadvantageous for detection.

\subsubsection{Summary}
\label{sssec:summary}
\begin{table}[htb]
\begin{ruledtabular}
\begin{tabular}{ccccc}
 & \multicolumn{2}{c}{$\alpha=0.1$} & \multicolumn{2}{c}{$\alpha=10^{-5}$} \\
$-\log(G\mu)$ & R\footnotemark[1] & NR\footnotemark[2] & R\footnotemark[1] & NR\footnotemark[2] \\
\hline
12 & 0 & $71\pm 8$ & 0\footnotemark[3] & $233\pm 16$ \\
14 & 0 & $1\pm 1$ & 0 & $4\pm 2$ \\
16 & 0 & $2\pm 1$ & 0 & 0\footnotemark[3] \\
17 & 0\footnotemark[3] & $50\pm 7$ & - & - \\
18 & $23\pm 5$ & $195\pm 15$ & 0\footnotemark[3] & $6\pm 2$ \\
19 & $47\pm 7$ & $47\pm 7$ & - & - \\
20 & $7\pm 3$ & $6\pm 3$ & 0\footnotemark[3] & 0\footnotemark[3] \\
\end{tabular}
\end{ruledtabular}
\footnotetext[1]{With the rocket effect}
\footnotetext[2]{Without the rocket effect}
\footnotetext[3]{Not all realizations in the simulation set have 0 events.}
\caption{Numbers of events for all 24 simulation sets of the toy model with $\Delta f$ corresponding to $T_{\mathrm{obs}}=1\ \mathrm{month}$.}
\label{tab:events}
\end{table}

We summarize in table \ref{tab:events} numbers of events for all 24 simulation sets of the toy model with $\Delta f$ corresponding to $T_{\mathrm{obs}}=1\ \mathrm{month}$. The third column shows that as expected, the rocket effect eliminates all events with $\alpha=10^{-5}$, including at $G\mu=10^{-12}$ which has a very high event number without the rocket effect. Recall from section \ref{sssec:clumptho} that assuming perfect clumping of loops in the Galaxy is an overestimation for $\alpha=10^{-5}$, implying that detection is actually even less likely than what our results suggest. This also means that detection is not possible for loops created from the small-loop paradigm. Our attention should be focused on big loops, $\alpha=0.1$, where the hope of detection rests upon. The viable region of the parameter space for LISA detection of harmonic signal from individual loops is therefore $G\mu\lesssim 10^{-18}$ and $\alpha=0.1$.

One more thing to note is that despite the fact that the 24 simulation sets above have $\Delta f$ corresponding to $T_{\mathrm{obs}}=1\ \mathrm{month}$, the guidance they provide on the interesting region of the parameter space for further analysis should be applicable to our goal of making predictions for a 1-year LISA mission. To see this, we first note that in the limit when $p_k$ is low, directly decreasing $\Delta f$ without rebinning loop abundance just increases the number of events correspondingly. This fact is verified by the statistical result from the simulation set with the rocket effect at $G\mu=10^{-18}$ and $\alpha=0.1$, in the alternative toy model performed with $\Delta f$ corresponding to $T_{\mathrm{obs}}=1\ \mathrm{year}$, where the number of events is
\begin{equation}
\label{eqn:numevent1y}
E_{\mathrm{D, 1y}}=283\pm 15,
\end{equation}
computed without incorporating higher harmonics, and should therefore be compared with eq. \ref{eqn:events1m}. Clearly, the difference is consistent with a factor of 12. Recall from the description of the toy models at the beginning of this section, rebinning loop abundance in the Galaxy for each frequency bin from the toy models into the physical model results in reductions by orders of magnitude as shown in eq. \ref{eqn:rebinfac}, especially at higher frequency where most events are found. From the discussion in section \ref{sssec:alpha}, for the simulation sets in the interesting region of the parameter space, the corresponding reduction in $p_k$ should be by a significant power-law of that factor which can be up to $\sim 10^{-7}$. This reduction more than offsets the enhancement shown above, and therefore results from the 24 sets in the toy model with $\Delta f$ corresponding to $T_{\mathrm{obs}}=1\ \mathrm{month}$ are already overestimations even for the physical simulation sets with $T_{\mathrm{obs}}=1\ \mathrm{year}$.

\subsubsection{Implication on Physical Results}
\label{sssec:impphys}
Arguments above already outline the basics for renormalizing results obtained from the toy models into those for the physical Galaxy. We now offer a concrete description of the renormalization procedure. The binning in $t_{\mathrm{c}}$ in eq. \ref{eqn:ininumdens} translates into a binning in $f$,
\begin{equation}
\label{eqn:fbinfull}
\left|\frac{d\ln f}{dt_{\mathrm{c}}}\right|=\left|\frac{d\ln l}{dt_{\mathrm{c}}}\right|=\frac{\alpha+\Gamma G\mu}{l}\simeq\frac{\alpha}{\alpha t_{\mathrm{c}}-\Gamma G\mu\left(t_0-t_{\mathrm{c}}\right)}.
\end{equation}
When loop decay is not significant,
\begin{equation}
\label{eqn:nodecay}
\alpha t_{\mathrm{c}}\gg\Gamma G\mu\left(t_0-t_{\mathrm{c}}\right),
\end{equation}
and eq. \ref{eqn:fbinfull} simplifies to the expected relation
\begin{equation}
\label{eqn:fbin}
\frac{\Delta f}{f}\simeq\frac{\Delta t_{\mathrm{c}}}{t_{\mathrm{c}}}=0.1.
\end{equation}
In the decay regime where the condition eq. \ref{eqn:nodecay} is not satisfied, natural frequency bins are no longer regular log bins, and can become enormous according to eq. \ref{eqn:fbinfull}, meaning that the already reduced population of loops in the Galaxy of the decay regime actually emits GW spanning a wide range of frequencies. The relation eq. \ref{eqn:fbin} is largely applicable to all simulation sets in the interesting region of the parameter space from previous discussions. Then, rebinning from the toy models into the physical model reduces the number of loops in the frequency bin by a bin reduction factor
\begin{equation}
\label{eqn:rebinfac}
\eta(f)=\frac{\Delta f_{\mathrm{obs}}}{\Delta f}=\frac{\frac{1}{T_{\mathrm{obs}}}}{0.1f}=\frac{10}{T_{\mathrm{obs}}f}\ll 1.
\end{equation}

Recall from section \ref{sssec:clm}, a given frequency bin $i$ in the toy models can register either no event or one event with some small probability $p_i$. This means that the total number of events for a simulation set can be interpreted as the summation of outcomes from a collection of Bernoulli trials (random trials with two possible outcomes, such as flipping a coin), with probabilities $p_i$ and $1-p_i$ for outcomes 1 and 0, respectively. Though the actual $p_i$ is unknown, from eqs. \ref{eqn:Pbin1} and \ref{eqn:Pbin2}, we know that $p_i$ is roughly reduced by a power-law of loop abundance which can be represented by the factor $\eta(f)^{\kappa}$ after rebinning, with the scaling parameter $\kappa=\frac{2}{3}$ when other sources of noise strongly dominate, and $\kappa=\frac{5}{6}$ when they only dominate weakly. From results of sets without the rocket effect and with $G\mu=10^{-18}$ for both values of $\alpha$, we see that the set with smaller $\alpha$ has a lower event number by a factor of $\sim 33$. Given that the difference in loop abundance is by about two orders of magnitude, we see that the actual simulation sets in fact contain a mixture of scenarios corresponding to different values of $\kappa$. We therefore compute predictions for physical detection probabilities for both cases, which should be interpreted as estimates of upper and lower bounds.

A set of $n$ Bernoulli trials follows the binomial distribution with the expectation value $np$. Adopting a frequentist interpretation, we can therefore effectively account for the rebinning without knowing $p_i$ by changing the success outcome for a frequency bin from 1 to $\eta(f)^{\kappa}$, effectively altering $n$. Then the predicted detection number for a simulation set in the physical model is
\begin{equation}
\label{eqn:numdet1y}
N_{\mathrm{P}}\sim\sum_i\eta\left(f_i\right)^{\kappa}=\langle\eta(f)^{\kappa}\rangle N_{\mathrm{T}},
\end{equation}
where the sum is over bins with events in the toy model. The more straightforward but less rigorous way of interpreting this relation is to think of $\eta\left(f_i\right)^{\kappa}$ as effective physical detection numbers for frequency bins.

Since events mostly occur at higher frequency, $\eta\left(f_i\right)$ is very small, and $N_{\mathrm{P}}$ is less than unity. It is therefore more suitable to interpret $N_{\mathrm{P}}$ as the total physical detection probability, $p_{\mathrm{det}}$, defined as the probability that at least one detection is made in the entire observed frequency range over the course of the mission. The more proper way of estimating $p_{\mathrm{det}}$ is
\begin{equation}
\label{eqn:probdet1y}
p_{\mathrm{det}}\sim 1-\prod_i\left(1-\eta\left(f_i\right)^{\kappa}\right).
\end{equation}
This interpretation with $\eta\left(f_i\right)^{\kappa}$ in the equation is well-defined in the limit $\eta\left(f_i\right)^{\kappa}\ll 1$ to the point that $N_{\mathrm{P}}$ is less than unity, because
\begin{equation}
\label{eqn:pdetjust}
p_{\mathrm{det}}\sim 1-\prod_i\left(1-\eta\left(f_i\right)^{\kappa}\right)\approx\sum_i\eta\left(f_i\right)^{\kappa}\sim N_{\mathrm{P}}
\end{equation}
after expansion and only keeping terms up to linear order in $\eta\left(f_i\right)^{\kappa}$. We should clarify that in writing down eq. \ref{eqn:probdet1y}, we are not taking $\eta\left(f_i\right)^{\kappa}$ to be the actual physical detection probability for the frequency bin, $p_{\mathrm{p},k}$. If the real $p_{\mathrm{p},k}$ is known, then the expression for $p_{\mathrm{det}}$ will be
\begin{equation}
p_{\mathrm{det}}=1-\prod_{k\in\mathrm{all\ bins}}\left(1-p_{\mathrm{p},k}\right),
\end{equation}
where the index $k$ is over all physical frequency bins. Clearly, $\eta\left(f_i\right)^{\kappa}\gg p_{\mathrm{p},k}$, as otherwise
\begin{equation}
N_{\mathrm{P}}=\sum_{k\in\mathrm{all\ bins}} p_{\mathrm{p},k}\gg 1
\end{equation}
from eq. \ref{eqn:rebinfac}, and the condition for eq. \ref{eqn:probdet1y} is not satisfied. In writing eq. \ref{eqn:probdet1y} with $\eta\left(f_i\right)^{\kappa}$, we are in fact implicitly estimating $p_{\mathrm{p},k}$ through the overall statistics of events in the toy model modulated by $\eta\left(f_i\right)^{\kappa}$ as a result of rebinning. The more intuitive and less rigorous way of interpreting eq. \ref{eqn:pdetjust} is simply that if the expected $N_{\mathrm{P}}$ is less than unity, say 0.1, then we reasonably expect to repeat the simulation set 10 times before a detection is made, i.e. the probability that a detection is made by performing the simulation set once can be said to be 0.1.

Eq. \ref{eqn:probdet1y} assumes that the $T_{\mathrm{obs}}$ used to bin the simulation set in the toy model is the actual LISA mission duration, and is therefore only applicable to the single simulation set in the alternative toy model. Rebinning results from the toy model with $\Delta f$ corresponding to $T_{\mathrm{obs}}=1\ \mathrm{month}$ requires some further manipulation. To obtain a correct estimate, we first compute $\eta(f)$ with $T_{\mathrm{obs}}=1\ \mathrm{year}$, producing the correct $\eta(f)$ for the physical model. The problem now is just a lack of frequency bins, i.e. a lack of Bernoulli trials. The solution is to effectively duplicate each frequency bin 12 times in eq. \ref{eqn:probdet1y}. This procedure produces reasonable estimates of $p_{\mathrm{det}}$, because loop abundance in the Galaxy is a smooth and slowly varying function of frequency, and therefore $p_{\mathrm{p},k}$ and $\eta(f_k)$ in nearby bins do not differ significantly. Then $p_{\mathrm{det}}$ can be estimated as
\begin{equation}
\label{eqn:probdet}
p_{\mathrm{det}}\sim 1-\prod_j\left(1-\eta\left(f_j\right)^{\kappa}\right)^{12},
\end{equation}
where we use a different index $j$ from that in eq. \ref{eqn:probdet1y} to indicate that the product is now over frequency bins with events in the toy model with $\Delta f$ corresponding to $T_{\mathrm{obs}}=1\ \mathrm{month}$.

\begin{figure}[htb]
\includegraphics[width=\columnwidth]{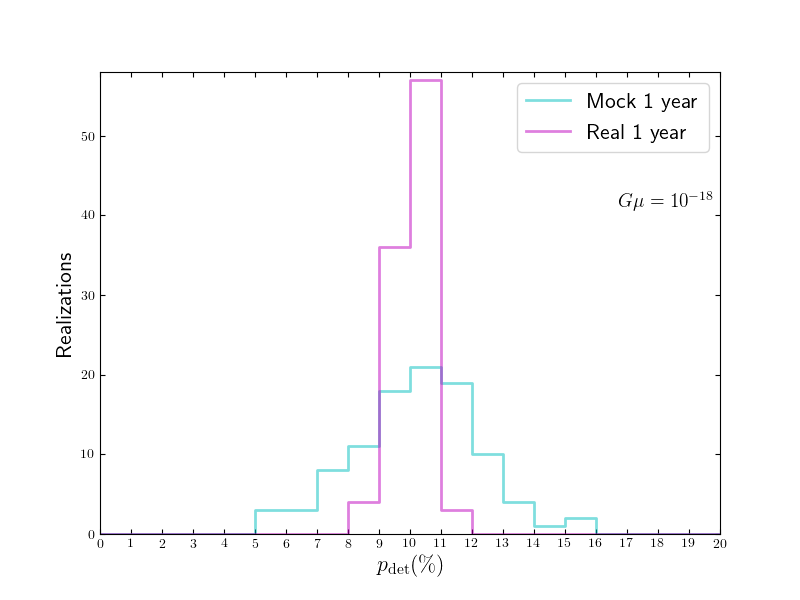}
\caption{Histograms of the predicted $p_{\mathrm{det}}$ with $\kappa=\frac{2}{3}$, from the simulation sets with $G\mu=10^{-18}$ and $\alpha=0.1$, for $T_{\mathrm{obs}}=1\ \mathrm{year}$. The histograms in cyan and magenta are computed using eqs. \ref{eqn:probdet} and \ref{eqn:probdet1y} from the toy models with $\Delta f$ corresponding to $T_{\mathrm{obs}}=1\ \mathrm{month}$ and $1\ \mathrm{year}$, respectively.}
\label{fig:rprobstats}
\end{figure}

We verify the derivations above by directly comparing results from eqs. \ref{eqn:probdet} and \ref{eqn:probdet1y}. We compute separately values of the predicted $p_{\mathrm{det}}$ with $\kappa=\frac{2}{3}$ from the simulation sets with $G\mu=10^{-18}$ and $\alpha=0.1$, in the toy models with $\Delta f$ corresponding to $T_{\mathrm{obs}}=1\ \mathrm{month}$ and $1\ \mathrm{year}$, respectively, with the histograms presented in fig. \ref{fig:rprobstats}. The same as in section \ref{sssec:summary}, GW emission is concentrated in the fundamental mode for the latter, and we use the same configuration for the former for consistency. Of course, this should not have any significant effects on the results. While these toy models have very different event numbers, as can be seen clearly by comparing eqs. \ref{eqn:events1m} and \ref{eqn:numevent1y}, the histograms of the predicted $p_{\mathrm{det}}$ are centered at similar values. Statistically, they basically produce the same prediction
\begin{eqnarray}
\label{eqn:probdetpf18}
p_{\mathrm{det,m}} & = & (10\pm 2)\%; \\
\label{eqn:probdetpfr18}
p_{\mathrm{det,r}} & = & (10.1\pm 0.6)\%.
\end{eqnarray}
Thus, we in fact can make predictions for the physical model from the toy model with a larger $\Delta f$, and a smaller $\Delta f$ simply leads to better statistics, which is not important for our purpose as we will discuss below.

Before we start making predictions, we first need to explain an important point to keep in mind when interpreting such predictions. We remarked earlier that the simulation set above actually contains a mixture of scenarios corresponding to different values of $\kappa$. GW emission by loops with this $G\mu$ is sufficiently strong, that at some frequencies, the Galactic confusion is already a significant source of noise. On the other hand, renormalizing into the physical model greatly reduces the Galactic confusion, meaning that here $\kappa=\frac{2}{3}$. Thus, in this case, $p_i$ first scales with $\kappa=\frac{5}{6}$ and then transitions to $\kappa=\frac{2}{3}$ during renormalization. Since the Galactic confusion is lower with lower $G\mu$, we expect the prediction made with $\kappa=\frac{2}{3}$ to be a better estimate at $G\mu=10^{-20}$.

\begin{figure}[htb]
\includegraphics[width=\columnwidth]{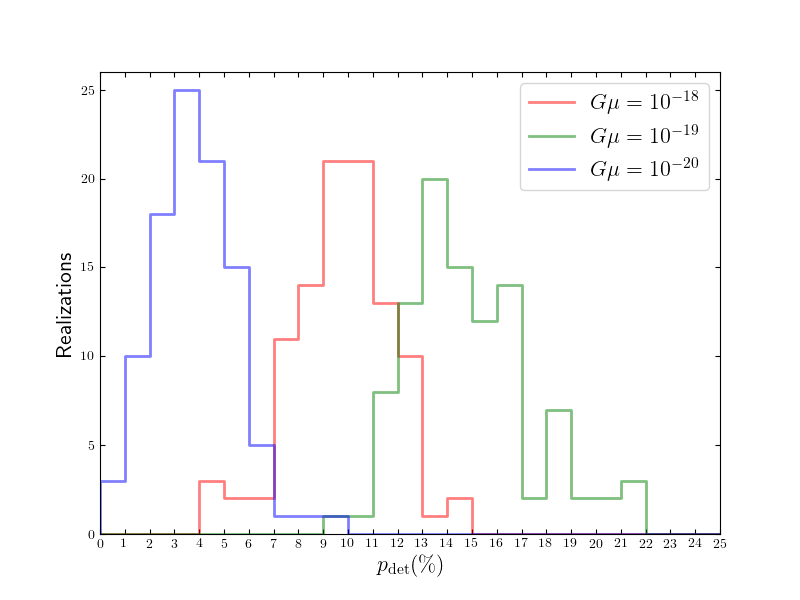}
\caption{Histograms of the predicted $p_{\mathrm{det}}$ with $\kappa=\frac{2}{3}$, from the simulation sets with $\alpha=0.1$, for $T_{\mathrm{obs}}=1\ \mathrm{year}$.}
\label{fig:rprobcomp}
\end{figure}

We are now able to make predictions for $p_{\mathrm{det}}$ for the interesting region of the parameter space, with the histograms of predictions with $\kappa=\frac{2}{3}$ presented in fig. \ref{fig:rprobcomp}. At $3\sigma$, the predicted bounds are
\begin{eqnarray}
\label{eqn:probdetp18}
0\% \lesssim & p_{\mathrm{det},18} & \lesssim 16\%; \\
\label{eqn:probdetp19}
0\% \lesssim & p_{\mathrm{det},19} & \lesssim 22\%; \\
\label{eqn:probdetp20}
0\% \lesssim & p_{\mathrm{det},20} & \lesssim 9\%.
\end{eqnarray}
We expect the upper bound in eq. \ref{eqn:probdetp20} to be a more decent estimate of the actual $p_{\mathrm{det}}$, while the other two are probably far too optimistic. The process of renormalization effectively converts simulation sets from the toy models into those in the physical model with more realizations, which is the underlying reason that allows us to make statistical predictions, and the statistics really comes as a result of sampling of subsets of these effective realizations. Of course, the conversion is far from accurate, and the predictions are only rough estimates on bounds. We will therefore still simulate the physical Galaxy despite the relatively unpromising predictions.

\section{Physical Results}
\subsection{Results for the Physical Model}
\label{sssec:physsetup}
\label{sssec:physresults}
With the obvious exception of cosmic string loop abundance in the Galaxy, the configuration and methodology for simulating the physical model are identical to those of the toy models. Equipped with the knowledge of the interesting region of the parameter space from the toy models, we run simulations with the rocket effect for $G\mu\in [10^{-18}, 10^{-19}, 10^{-20}]$ and $\alpha=0.1$, all with $T_{\mathrm{obs}}=1\ \mathrm{year}$. The power of GW emission from loops is concentrated in the fundamental mode in all physical simulation sets to conserve computing resources, as we have seen from section \ref{sssec:GWspec} that this makes little difference, and any difference is towards a very slight overestimation of detection.

\begin{figure}[htb]
\includegraphics[width=\columnwidth]{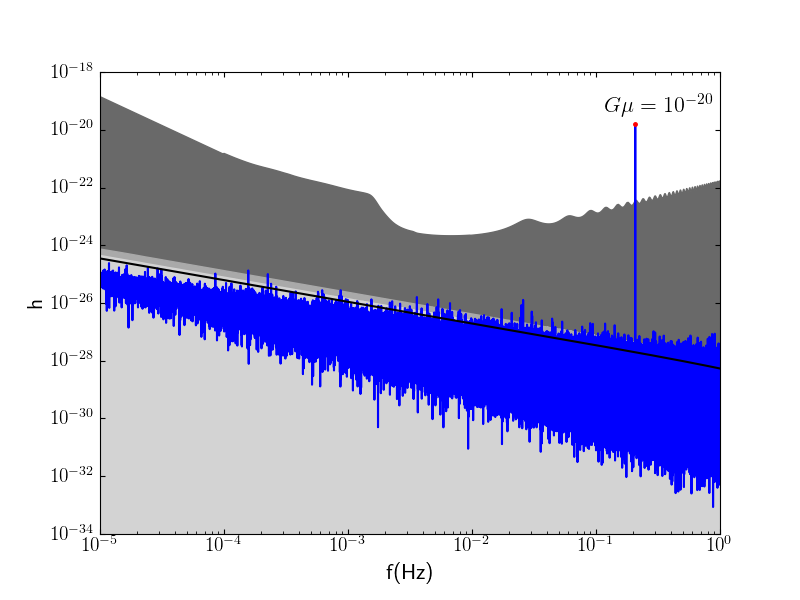}
\caption{Snapshot of a realization capturing a detection with $\varrho\sim 450$ in the simulation set in the physical model with the rocket effect at $G\mu=10^{-20}$ and $\alpha=0.1$. Plot composition is identical to that of fig. \ref{fig:decetcomp}.}
\label{fig:physdet20}
\end{figure}

Even though loop abundance in the physical model is severely reduced, by up to a factor of $\sim 3\times 10^{-7}$, there still are detections, with an example at $G\mu=10^{-20}$ presented in fig. \ref{fig:physdet20}. Even though the Galactic background is simply drowning in the LISA instrumental noise with the Galactic WD binaries background given the reduced loop abundance, there is nevertheless, a loop which happens to be located extremely closely, to the point that its harmonic GW emission greatly overwhelms everything else. This detection has $\varrho\sim 450$, making it unmissable as a persistent source. Another feature to note is that here, the Galactic background is approximately comparable to the stochastic background, which should be general for the physical Galaxy regardless of parameters as long as loops in the Galaxy remain well clumped.

\begin{figure}[htb]
\includegraphics[width=\columnwidth]{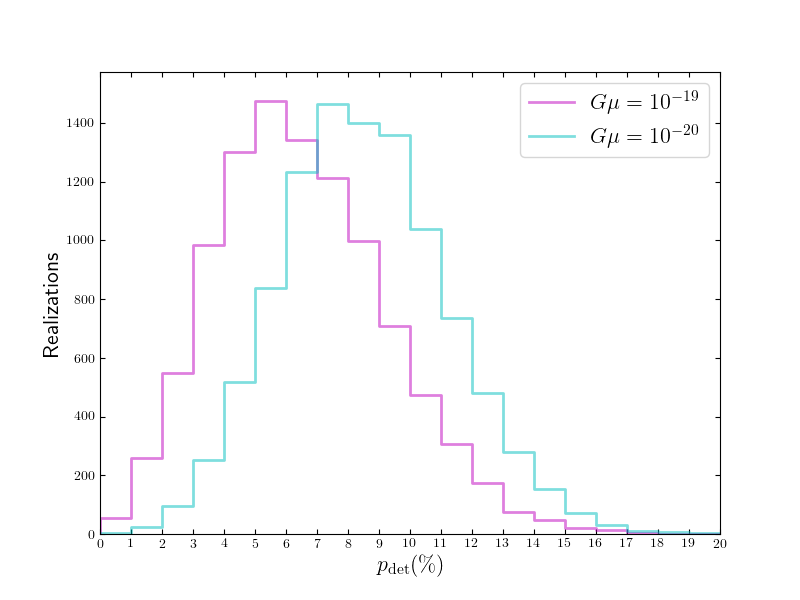}
\caption{Histograms from bootstrapping the numbers of detections from realizations in the simulation sets in the physical model with the rocket effect at $G\mu=10^{-19}$ (magenta) and $10^{-20}$ (cyan), with $\alpha=0.1$.}
\label{fig:histoboot}
\end{figure}

The overall statistical results however, do not look as optimistic. First of all, we have no detection at $G\mu=10^{-18}$. The detection numbers increase for lower $G\mu$, but remain low compared to the number of realizations in a set. By a similar reasoning as that which led to eq. \ref{eqn:pdetjust}, we can interpret these results as probabilities of detection over the course of the mission, $p_{\mathrm{det}}$. This interpretation is exact for the set at $G\mu=10^{-20}$, where no realization contains more than one detection. We obtain estimates of the detection statistics for $G\mu=10^{-19}$ and $10^{-20}$ by bootstrapping the detection numbers from realizations in the simulation sets with $10^4$ samples, with the histograms presented in fig. \ref{fig:histoboot}. We verified that similar statistics can be obtained with $10^3$ samples, meaning that the bootstrapping process has fully converged.

\begin{table}[htb]
\begin{ruledtabular}
\begin{tabular}{cc}
$-\log(G\mu)$ & $p_{\mathrm{det}}$ \\
\hline
$\lesssim 18$ & $<1\%$ \\
19 & $(6\pm 3)\%$ \\
20 & $(8\pm 3)\%$ \\
21\footnotemark[1] & $(6\pm 2)\%$ \\
22\footnotemark[1] & $(1\pm 1)\%$ \\
$\gtrsim 23$\footnotemark[1] & $<1\%$ \\
\end{tabular}
\end{ruledtabular}
\footnotetext[1]{Estimated from the simulation set at $G\mu=10^{-20}$.}
\caption{Detection probabilities in the physical model with $\alpha=0.1$.}
\label{tab:detections}
\end{table}

Further decreasing $G\mu$ beyond $10^{-20}$ provides no benefit to enhancing detection, because loop abundance is not significantly enhanced, as loops in the LISA frequency range are fully out of the rocket effect and the decay regimes, while GW emission from loops weakens. This enables us to estimate $p_{\mathrm{det}}$ for $G\mu<10^{-20}$ from the set at $G\mu=10^{-20}$ by effectively increasing the SNR requirement, with the results summarized in table \ref{tab:detections}. Loops with $G\mu\sim 10^{-19}$ - $10^{-21}$ have GW emission that is just sufficiently weak to keep them outside of the rocket effect and the decay regimes, and therefore this region represents the sweet spot in the parameter space for detection. These results confirm our prediction from section \ref{sssec:impphys}, that detection is not very likely even under the most favorable conditions.

\subsection{Results with Massless Neutrinos}
\label{ssec:masslessneu}
\label{sssec:masslesshar}
\label{sssec:masslessbg}
We present here results with the standard $\Lambda$CDM cosmology with massless neutrinos and Planck 2018 cosmological parameters\cite{Planck2018,Auclair2019},
\begin{eqnarray}
H_0 & = & 67.8\ \mathrm{km/s/Mpc}; \\
\Omega_{\mathrm{m},0} & = & 0.308; \\
\Omega_{\mathrm{r},0} & = & 9.1476\times 10^{-5}; \\
\Omega_{\mathrm{\Lambda},0} & = & 1-\Omega_{\mathrm{m},0}-\Omega_{\mathrm{r},0}.
\end{eqnarray}
Compared to our treatment of the neutrino mass in the toy models described in section \ref{sssec:params}, this further reduces loop abundance by about a factor of $\sim 10$. Given the already low $p_{\mathrm{det}}$ summarized in table \ref{tab:detections}, this exercise is undertaken chiefly in the spirit of completeness and for ease of comparison to other studies. An estimate can be obtained by using the heuristic method developed in section \ref{sssec:impphys}. From eq. \ref{eqn:probdet1y} with $\kappa=\frac{2}{3}$ and table \ref{tab:detections}, we expect $\sim 8/10^{\frac{2}{3}}\approx 2$ detections out of the total of 100 realizations for the simulation set with $G\mu=10^{-20}$, and up to $\sim 1$ detection for the set with $G\mu=10^{-19}$.

Actual results from the physical simulation sets with the standard cosmology above are in good agreement with expectation. We have no detection in the set with $G\mu=10^{-19}$, and indeed two detections in the set with $G\mu=10^{-20}$. Thus, we can conclude that direct detection of harmonic GW from individual loops by LISA is very unlikely for field theory cosmic strings, with
\begin{equation}
p_{\mathrm{det}}\lesssim 2\%.
\end{equation}

\begin{figure}[htb]
\includegraphics[width=\columnwidth]{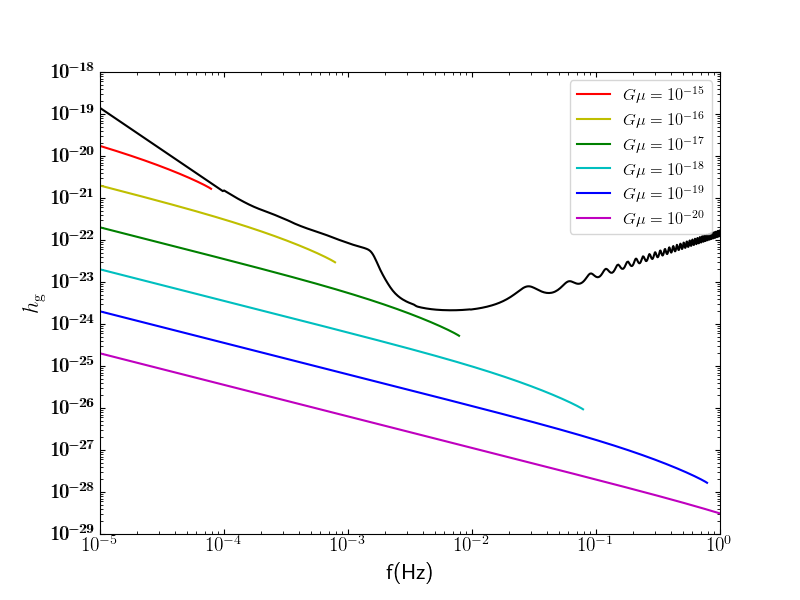}
\caption{The physical Galactic background with the $\Lambda$CDM cosmology and massless neutrinos, plotted in colors following the rainbow order from $G\mu=10^{-15}$ to $10^{-20}$ with $\alpha=0.1$, and compared to the LISA instrumental noise with the Galactic WD binaries background (black).}
\label{fig:pgalhml}
\end{figure}

For detecting the unresolved Galactic background, we present in fig. \ref{fig:pgalhml}, the physical Galactic background with the $\Lambda$CDM cosmology and massless neutrinos, binned for $T_{\mathrm{obs}}=1\ \mathrm{year}$, and can be compared directly to the LISA instrumental noise. Clearly, the first impression we have is that the Galactic background is not detectable. Curves are terminated at $f_{\odot}$ when loops in the Galaxy enter the decay regime, where eq. \ref{eqn:fbin} is no longer valid and should be replaced by eq. \ref{eqn:fbinfull}. We can see from the latter that as loops enter the decay regime, the denominator quickly becomes very small, and the number density in a LISA frequency bin decreases rapidly. This means that loops with similar $t_{\mathrm{c}}$ which, without significant loop decay should be in the same bin, are now stretched across many bins because their current sizes differ significantly, effectively greatly reducing loop abundance. Physically, this is a manifestation of loop evaporation. Thus, the Galactic background just drops rapidly in the decay regime, and there is no meaning in calculating it there for our purpose, with the unresolved background itself becoming undefined. For all practical purposes, the physical Galactic background basically vanishes quickly beyond $f_{\odot}$. Loops with $G\mu>10^{-15}$ in the LISA frequency range are all in the decay regime. Even if we ignored loop decay and insisted on using eq. \ref{eqn:fbin}, the change in slope due to the rocket effect at $f_{\odot}$ means that the Galactic background would still not be detectable.

Recall from section \ref{sssec:hLISA} that the low-frequency part of the sensitivity curve below $f\sim 10^{-3}\ \mathrm{Hz}$ is primarily due to the Galactic WD binaries background, and therefore the Galactic background with higher $G\mu$ is simply not detectable regardless of the detector sensitivity due to confusion with Galactic WD binaries. Comparing to the stochastic background in fig. \ref{fig:hnu} reduced by a factor of $\sim 3$ to account for cosmology, we see that they are about comparable outside the decay regime, the same conclusion we made earlier by observing fig. \ref{fig:physdet20}. However, the stochastic background is easier to detect since we also receive GW emitted by loops from the past, greatly enhancing signal in the decay regime. This is probably what makes the stochastic background detectable by LISA for $G\mu\gtrsim 10^{-17}$\cite{Auclair2019}. We are therefore led to the conclusion that despite the significant clumping of loops in the Galaxy, detection of Galactic loops is likely only possible for very strong signal from individual loops, whether harmonic signal from extremely nearby loops for lower $G\mu$ analyzed in this work, or burst signal from background loops for higher $G\mu$ analyzed in ref. \cite{ChernoffBurst}.

\section{Conclusion}
\label{sec:conclude}
We developed through this study, a robust framework for numerically simulating cosmic string loops clumped in the Galaxy taking into account effects of loop recoil as a result of the overall anisotropy in their GW emission, and generating the expected signal observable by LISA for a 1-year mission. The methodology we developed to accomplish this task is highly flexible, and can be adapted relatively easily to produce results from simulations configured with any reasonable loop formation and cosmological models. Moreover, the toolset in our simulations represents particularly efficient methods for simulating loops in the Galaxy for predicting detection of harmonic GW from resolved loops located in proximity to the solar system, even with extremely high loop abundances such as those of the toy models. Along the way, we also derived a method for estimating detection rates when loop abundance changes.

Compared to the power and adaptability of the simulation methodology developed to accomplish our objective, results obtained from the simulation sets of the physical Galaxy are arguably less exciting. The immediate conclusion we can draw based on the results summarized in table \ref{tab:detections}, as well as the results from section \ref{sssec:masslesshar}, is that detection of harmonic GW from resolved loops by LISA is unlikely for field theory cosmic strings anywhere in the parameter space, even under the most favorable conditions. We further conclude from fig. \ref{fig:pgalhml} that LISA will also have difficulty detecting the unresolved Galactic background, despite the enhancement due to complete clumping of loops in the Galaxy.

In our approach of directly resolving harmonic signal from individual loops, the critical ingredient for detection is the possibility of them being located in extreme proximity. This requirement essentially eliminates higher $G\mu$ due to decreased loop abundance as a result of loop decay, and even more importantly, because of the rocket effect which completely prevents detection when $r_{\mathrm{tr}}<r_{\odot}$. Our results show that resolving harmonic signal from individual loops is optimized when $G\mu$ is just sufficiently low to keep loops outside of the decay and the rocket effect regimes, $10^{-21}\lesssim G\mu\lesssim 10^{-19}$ for the LISA frequency range. This should be contrasted to the other approach focusing on detecting bursts from loops, which instead prefers $G\mu>10^{-15}$\cite{ChernoffBurst}. This is because for a given frequency range, the target loops for burst detection are larger and younger, and therefore are not significantly affected by loop decay and the rocket effect even with higher $G\mu$. Both approaches prefer loops with large initial sizes, $\alpha=0.1$, because of higher initial number densities, and more importantly, better loop clumping in the Galaxy.

We mentioned in section \ref{sssec:rockettho} that the rocket effect incorporated into our simulation is derived\cite{CH09} with the cusps-dominated GW emission spectrum, associated with the highest degree of overall anisotropy in the direction of emission, maximizing the recoil experienced by loops. The overall direction of this recoil is additionally taken to remain constant over time. In reality, loops probably encompass a wide range of trajectories with various small-scale features, with the overall directions of anisotropy in their GW emissions quite possibly changing over time. Thus, the treatment of the rocket effect incorporated is a relatively strong version\cite{ChernoffBurst}. For the solar system orbit however, the possibility of the rocket effect being somewhat weaker in reality does not have a significant effect on our results, because here, the rocket effect essentially tracks the decay regime for our purpose as we discussed in section \ref{sssec:rockettho}, where loop abundance drops quickly as we discussed in section \ref{sssec:masslessbg}, making this frequency range unimportant for detection regardless of the rocket effect. We also already adopt complete clumping of Galactic loops in our simulation, which cannot be further enhanced by a potentially weaker version of the rocket effect.

An important avenue for future exploration is provided by string theory, where the richness of the theory allows for greatly relaxed theoretical restrictions on the properties of cosmic superstrings\cite{Chernoff2014}. An important such property is $P_{\mathrm{int}}$, which is now permitted to be much less than 1, down to the very low $P_{\mathrm{int}}\sim 10^{-3}$\cite{lowPint,Chernoff2014}. It may be somewhat counterintuitive that as a result of the scaling solution of loop formation, a reduction in $P_{\mathrm{int}}$ actually leads to an enhancement in loop number density\cite{Hogan2006}. The amount of enhancement varies depending on the details of loop formation models\cite{Pint1,Pint2,Pint3,ChernoffBurst}, and can be as high as $P_{\mathrm{int}}^{-2}$\cite{Pint1,ChernoffBurst}, resulting in an enhancement by a factor of $10^6$. From our method for estimating detection rates eq. \ref{eqn:probdet1y} and the results in table \ref{tab:detections}, we see that taking into account cosmological models, such an enhancement means that direct detection of harmonic GW from individual loops will basically be a certainty for loops outside of the rocket effect and the decay regimes, $G\mu\lesssim 10^{-15}$, down to at least $G\mu\sim 10^{-22}$, with the Galactic background itself likely detectable down to $G\mu\sim 10^{-20}$. Even for a conservative enhancement by a factor of 100 with $P_{\mathrm{int}}^{-\frac{2}{3}}$\cite{Pint3,ChernoffBurst}, our estimate indicates close to $p_{\mathrm{det}}\sim 50\%$ by LISA for $G\mu\sim 10^{-20}$. The richness of the string theory framework leaves the door to direct detection of harmonic GW from loops wide open. Future studies are needed to explore the possibility of detecting such signals with various models of cosmic superstrings. LISA observations in search of harmonic signals from cosmic string loops will place combined constraints on the $(G\mu, P_{\mathrm{int}})$ parameter space, and have the potential to be experimental tests of string theory.

\begin{acknowledgments}
Simulations of Galactic cosmic string loops in this work were performed on the Midway High-Performance Computing Cluster, as part of the Cluster Partnership Program supported by the Kavli Institute for Cosmological Physics (KICP) at the University of Chicago.
\end{acknowledgments}

\bibliography{refs}

\end{document}